\documentclass{pasa}
\usepackage{graphicx}
\usepackage{amsmath}
\usepackage{gensymb} 
\usepackage{xfrac}
\usepackage[flushleft]{threeparttable}
\usepackage{amssymb}
\usepackage{bm}
\usepackage{multirow}

\usepackage{aas-macros}
\usepackage{hyperref}
\hypersetup{colorlinks,citecolor=blue,linkcolor=blue,urlcolor=blue}
\usepackage{xcolor}
\usepackage{acronym}
\usepackage{placeins}

\let\Gamma\varGamma
\let\Delta\varDelta
\let\Theta\varTheta
\let\Xi\varXi
\let\Pi\varPi
\let\Sigma\varSigma
\let\Upsilon\varUpsilon
\let\Phi\varPhi
\let\Psi\varPsi
\let\Omega\varOmega

\newcommand{\de}{\mathrm{d}}

\newcommand{\oc}{\Omega_{{\rm c},0}}
\newcommand{\om}{\Omega_{{\rm m},0}}
\newcommand{\ob}{\Omega_{{\rm b},0}}

\usepackage{comment}

\usepackage[nameinlink,noabbrev]{cleveref}
\Crefname{equation}{Eq.}{Eqs.}
\Crefname{section}{Sect.}{Sects.}
\Crefname{figure}{Fig.}{Figs.}
\crefname{equation}{Equation}{Equations}
\crefname{section}{Section}{Sections}
\crefname{figure}{Figure}{Figures}

\title[Cross-correlating EMU PS1 with CMB lensing]{Cross-correlating the EMU Pilot Survey 1 with
CMB lensing: Constraints on cosmology and galaxy bias with harmonic-space power spectra}
\author[K.\ Tanidis et al.]{K.\ Tanidis,$^{1}$\thanks{E-mail: konstantinos.tanidis@physics.ox.ac.uk}
J.\ Asorey,$^{2,3}$
C.S.\ Saraf,$^{4}$
C.L.\ Hale,$^{1}$
B.\ Bahr-Kalus,$^{5,6,7}$
D.\ Parkinson,$^{4}$
S.\ Camera,$^{6,7,5}$
R.P.\ Norris,$^{8,9}$
A.\ M. Hopkins,$^{10}$
M.\ Bilicki,$^{11}$ and
N.\ Gupta$^{12}$
\\
\affil{$^{1}$Department of Physics, University of Oxford, Denys Wilkinson Building, Keble Road, Oxford OX1 3RH, UK}
\affil{$^{2}$Departamento de F\'isica Te\'orica and IPARCOS, Universidad Complutense de Madrid, 28040 Madrid, Spain}
\affil{$^{3}$Departamento de F\'isica Te\'orica, Centro de Astropart\'iculas y F\'isica de Altas Energ\'ias, Universidad de Zaragoza, 50009 Zaragoza, Spain}
\affil{$^{4}$Korea Astronomy and Space Science Institute, 776 Daedeok-daero, Yuseong-gu, Daejeon 34055, Republic of Korea}
\affil{$^{5}$INAF -- Istituto Nazionale di Astrofisica, Osservatorio Astrofisico di Torino, Via Osservatorio 20, 10025 Pino Torinese, Italy}
\affil{$^{6}$Dipartimento di Fisica, Universit\`a degli Studi di Torino, Via P.\ Giuria 1, 10125 Torino, Italy}
\affil{$^{7}$INFN -- Istituto Nazionale di Fisica Nucleare, Sezione di Torino, Via P.\ Giuria 1, 10125 Torino, Italy}
\affil{$^{8}$Western Sydney University, Locked Bag 1797, Penrith, NSW 2751, Australia}
\affil{$^{9}$CSIRO Space \& Astronomy, P.O. Box 76, Epping, NSW 1710, Australia}
\affil{$^{10}$School of Mathematical and Physical Sciences, 12 Wally’s Walk Macquarie University, NSW 2109, Australia}
\affil{$^{11}$Center for Theoretical Physics, Polish Academy of Sciences, al. Lotników 32/46, 02-668 Warsaw, Poland }
\affil{$^{12}$CSIRO Space \& Astronomy, PO Box 1130, Bentley WA 6102, Australia }
}
\jid{PASA}
\doi{10.1017/pas.\the\year.xxx}
\jyear{\the\year}

\begin{document}
\begin{frontmatter}
\label{firstpage}
\maketitle

\begin{abstract}
We measured the harmonic-space power spectrum of galaxy clustering auto-correlation from the Evolutionary Map of the Universe Pilot Survey 1 data (EMU PS1) and its cross-correlation with the lensing convergence map of cosmic microwave background (CMB) from \textit{Planck} Public Release 4 at the linear scale range from $\ell=2$ to 500. We applied two flux density cuts at $0.18$ and $0.4$mJy on the radio galaxies observed at 944MHz and considered two source detection algorithms. We found the auto-correlation measurements from the two algorithms at the 0.18mJy cut to deviate for $\ell\gtrsim250$ due to the different criteria assumed on the source detection and decided to ignore data above this scale. We report a cross-correlation detection of EMU PS1 with CMB lensing at $\sim$5.5$\sigma$, irrespective of flux density cut. In our theoretical modelling we considered the \texttt{SKADS} and \texttt{T-RECS} redshift distribution simulation models that yield consistent results, a linear and a non-linear matter power spectrum, and two linear galaxy bias models. That is a constant redshift-independent galaxy bias $b(z)=b_g$ and a constant amplitude galaxy bias $b(z)=b_g/D(z)$. By fixing a cosmology model and considering a non-linear matter power spectrum with \texttt{SKADS}, we measured a constant galaxy bias at $0.18$mJy ($0.4$mJy) with $b_g=2.32^{+0.41}_{-0.33}$ ($2.18^{+0.17}_{-0.25}$) and a constant amplitude bias with $b_g=1.72^{+0.31}_{-0.21}$ ($1.78^{+0.22}_{-0.15}$). When $\sigma_8$ is a free parameter for the same models at $0.18$mJy ($0.4$mJy) with the constant model we found $\sigma_8=0.68^{+0.16}_{-0.14}$ ($0.82\pm0.10$), while with the constant amplitude model we measured $\sigma_8=0.61^{+0.18}_{-0.20}$ ($0.78^{+0.11}_{-0.09}$), respectively. Our results agree at $1\sigma$ with the measurements from \textit{Planck} CMB and the weak lensing surveys  and also show the potential of cosmology studies with future radio continuum survey data.
\end{abstract}

\end{frontmatter}


\section{Introduction}
\label{sec:intro}

A primary goal of large-scale structure experiments probing the late Universe, is to provide answers on the history of the growth of cosmic structures and also discover the nature of the unknown components that dominate in the Universe leading it to its recent accelerated expansion \citep[{e.g.}][]{Huterer_2023}. To achieve this, the tracers we choose should be able, on one hand, to cover a large patch of the observed sky, accessing this way both large and small cosmological scales and, on the other hand, to be deep enough so that we can reconstruct the growth of structure history as a function of time. However, these probes alone, are not able to address these aspects simultaneously. For instance, probes like weak gravitational lensing on galaxies, which is the effect of the distortions of galaxy shapes caused by the underlying matter field between us and the galaxies, or on the cosmic microwave background (CMB) \citep{Bartelmann_2001}, is an unbiased tracer of the matter field in the Universe. Nonetheless, it provides poor information on the redshift evolution of the galaxies and also has lower statistical power compared to the other large-scale structure probe, called galaxy clustering. This probe, though, is a biased tracer of the total matter field and the modelling needed to connect the two has been proven to be quite complex \citep{Kaiser1987, 10.1093/mnras/stw2443, Abbott:2018ydy, Desjacques_2018}. One way to overcome this and reconstruct the growth of structures, is to use redshift-space distortions in case there are accurate redshift estimates which are obtained spectroscopically \citep{Guzzo_2008, Blake_2013, Howlett_2015, Pezzotta_2017, Alam_2021}. Another way to overcome the limitations from the individual experiments, is to combine weak lensing and galaxy clustering data measurements \citep{Hu_2002, de_la_Torre_2017, Peacock_2018, Wilson_2019, Heymans_2021, White_2022, Garc_a_Garc_a_2021, Alonso_2023}. In addition, this multi-tracing increases the statistical power by accessing as much information as possible in the different cosmological scales as well as in redshift. 

In this framework, there has been a growing interest in deep radio continuum galaxy surveys. These surveys have the ability to scan enormous patches of the sky thanks to the large field of view of modern radio interferometers operating at low frequencies. There has been a variety of forecasting analyses in the literature arguing for their cosmological potential using the Square Kilometer Array \citep[hereafter SKAO][]{Raccanelli_2012, Jarvis_2015, Maartens2015, Bacon_2020} and also the benefit reaped when different radio populations are combined in a multi-tracer approach. In particular, several ultra-large scale effects can be detected with multi-tracing such as relativistic effects and the primordial non-Gaussianity \citep{Ferramacho_2014, Alonso_2015, Fonseca_2015, Bengaly_2019, Gomes_2019}.

When observing the Universe at frequencies between 0.1-10 GHz, wavelengths larger than those in optical and infrared, the main radio continuum emission mechanism is synchrotron radiation\footnote{Although, at 5-10GHz free-free emission starts to be also important.} \citep[{e.g.}][]{Condon_1992}. This is caused by relativistic electrons as they spiral in the magnetic fields. For this reason, the dominant populations of the radio galaxies are active galactic nuclei (hereafter AGNs), and star forming galaxies (hereafter SFGs). Regarding AGNs, there is a variety in the origin of sources as well as in their classifications. This includes the accretion mechanism of infalling material into central supermassive black holes \citep[{e.g.}][]{Best_2012, Heckman_2014}, AGN orientation with respect to the observer \citep{Antonucci_1993, Urry_1995} and also their morphology \citep[{e.g. } Type I \& II,][]{Fanaroff_1974}. As for SFGs, these are mainly spiral galaxies and they fall into two main categories. First is starburst galaxies, in which intensive star formation is present (star formation rate $\gtrsim 100 \text{ M}_\odot \text{yr}^{-1}$). The other category is normal star forming galaxies (star formation rate $\lesssim 100 \text{ M}_\odot \text{yr}^{-1}$) {\citep[e.g.][]{Wynn-Williams1986}}. One of the main advantages of observations at these frequencies is that dust contamination is negligible in the line-of-sight direction as well as in the intergalactic medium due to the long wavelengths at radio frequencies. This is especially relevant for SFG studies where their radio emission is an unbiased probe of the star formation rate \citep[{e.g.}][]{Bell_2003, Davies_2016, G_rkan_2018}.

There have been a number of past large-area radio continuum experiments like the NRAO VLA Sky Survey \citep[NVSS at 1.4GHz,][]{Condon_1998, Hotan_2021}, the TIFR GMRT Sky Survey \citep[TGSS-ADR at 150MHz,][]{Intema_2017} and the Sydney University Monongolo Sky Survey \citep[SUMSS,][]{Mauch_2003}. However, the current generation of radio surveys like the Australian Square Kilometre Array Pathfinder \citep[hereafter ASKAP,][]{johnston2007}, the Meer Karoo Array Telescope \citep[hereafter MeerKAT,][]{Jonas2009} and the LOw Frequency ARray \citep[hereafter LOFAR,][]{van_Haarlem_2013}, all of them precursors of SKAO, make an advance through much deeper observations together with the large sky coverage. In particular, ASKAP has a field of view of $\sim 30 \text{ deg}^2$ operating at 700-1800MHz thanks to its phased array feeds. LOFAR, similarly, has a field of view of $\sim 30 \text{ deg}^2$ at 150MHz, while MeerKAT has a field of view of $\sim 1 \text{ deg}^2$ at 1.2GHz. With these large fields of view achieved with radio interferometers, large and contiguous patches of the sky can be observed, accessing in this way large-scale structure information at very large scales (angular separations). 

In this work we use the Pilot Survey 1 of the Evolutionary Map of the Universe \citep[hereafter EMU PS1,][]{norris2011, EMUPS1} which uses ASKAP at 944MHz, covering a contiguous patch of $\sim 270 \text{ deg}^2$ at a depth of 25-30 $\mu \text{Jy/beam rms (root mean square)}$ and with a spatial resolution of 11-18 $\text{arcsec}$. By the end of its operation, EMU will cover the whole of the southern sky. 

As already mentioned, the radio continuum emission mechanism is synchrotron radiation, whose spectrum {typically lacks strong emission or absorption lines which} renders redshift measurements impossible\footnote{Technically, some sources can have spectral lines and the issue is that any potential frequency information is collapsed down.}. This results in large uncertainties on the redshift distribution of the galaxy sample and its properties, like the mass of host halos and galaxy bias. To shed light on radio sources' clustering properties, one solution is to cross-match with optical sources \citep[{e.g.}][]{Lindsay_2014,Hale18, Mazumder_2022}. 

Radio continuum sources overlap in redshift with the CMB lensing convergence field. This probe is sensitive to inhomogeneities of the matter distribution at high redshifts (peaking at $z\sim2$) and at comparable large volumes, making it ideal for cross-correlations with radio galaxies \citep[{e.g.}][]{Ade2013, Allison_2015} and also in the context of de-lensing studies \citep{Namikawa_2016}. Previous works on cross-correlation of radio galaxies with CMB lensing include \cite{Smith_2007}, where this combination was used to make the first CMB lensing detection and also \cite{Allison_2015} and \cite{Piccirilli22} to infer the galaxy bias of radio galaxies. Furthermore, the first and second data releases of the LOFAR Two-metre Sky Survey (LoTSS; \citealt{Shimwell_2019, Shimwell_2022}) radio catalogues were cross-correlated with CMB lensing from the \textit{Planck} satellite \citep{Planck2020}, in order to constrain the redshift distribution and the galaxy bias of the sample \citep{Alonso2021, Nakoneczny24}. These works have also shown that this cross-correlation can lift the degeneracy between the galaxy bias and the amplitude of the matter fluctuations. Here, we explore the auto-correlation and the cross-correlation of EMU PS1 with the latest CMB lensing convergence data (PR4) from \textit{Planck} \citep{Carron_2022} to place constraints on the galaxy bias of the sample and on the matter fluctuations amplitude and leave the redshift distribution parameterisation of radio sources with the help of optical surveys for a future work.

The paper is structured as follows: In \autoref{sec:theory} we describe the theoretical observables we use in our modelling. Then, in \autoref{sec:dataset} we present the data we use in our analysis. In \autoref{sec:methodology} we introduce the method used to construct the auto-correlation and cross-correlation measurements from the data, and also discuss the models and the error estimates we assume for our statistical analysis. The main results concerning the detection significance and the constraints on the galaxy bias and cosmology are shown in \autoref{sec:results}. Finally, we discuss our conclusions in \autoref{sec:conclusions}.

\section{Theory}
\label{sec:theory}


The harmonic-space power spectrum signal $S_{\ell}^{XY}$ between the projected quantities $X$ and $Y$, can be defined as 
\begin{equation}
    \left\langle X_{\ell m}\,Y^\ast_{\ell' m'}\right\rangle=S^{XY}_\ell\,\delta^{\rm K}_{\ell\ell'}\,\delta^{\rm K}_{m m'}\;,
    \label{eq:Harmonic_definition}
\end{equation}
where $X_{\ell m}$ and $Y_{\ell m}$ denote the coefficients of the harmonic expansion for the statistically isotropic fields of interest $X$ and $Y$, while $\delta^{\rm K}$ is the Kronecker symbol. In this work we focus on the fluctuations of the galaxy number counts $\delta_g$ and the convergence field $\kappa$. Both are later discussed in detail in \autoref{subsec:GC} and \autoref{subsec:CMBLens}, respectively.

For broad redshift distributions, as is the case in radio continuum surveys \citep[{e.g.}][]{Tanidis_mag}, the harmonic-space power spectrum between the two quantities $X$ and $Y$ can be written in the Limber approximation \citep{Limber1953,1992ApJ...388..272K} as
%
\begin{multline}
S^{XY}_{\ell, \text{th}}=\int_0^{\chi_h} \frac{\de \chi}{\chi^2}\, W^X(\ell,\chi)\,W^Y(\ell, \chi)\, \\
\times P_{mm}\left(k=\frac{\ell+1/2}{\chi},\chi\right)\;,
    \label{eq:Harmonic_limber}
\end{multline}
where $\chi(z)$ is the comoving distance at a given redshift $z$ for flat cosmologies, $\chi_h$ the co-moving distance at the horizon, $P_{mm}$ is the matter power spectrum and $k=|\vec k|$ with $\vec k$ the wave vector. We use the notation $S_\text{th}$ for the model harmonic-space spectrum to distinguish it from $S$ which is the measured harmonic-space spectrum signal from definition in \autoref{eq:Harmonic_definition}. The general redshift and scale dependent kernel $W^X(\ell,\chi)$ can take different expressions depending on the desired observable. These observables are described in \autoref{subsec:GC} and \autoref{subsec:CMBLens}.

\subsection{Galaxy Clustering}
\label{subsec:GC}


%
Galaxies are well-known biased tracers of the dark matter field \citep{Kaiser1987}. In general, this bias can be considered to be redshift and scale dependent. Assuming Gaussian initial curvature perturbations, this scale dependence is especially relevant at non-linear scales, where the bias is non-local \citep[{e.g.}][]{10.1093/mnras/stw2443, Desjacques_2018}. Nevertheless, at sufficiently large scales as we probe here $(k~\lesssim 0.2 \,h\,\mathrm{Mpc}^{-1})$, we can assume that it is only redshift dependent \citep[{e.g.}][]{Abbott:2018ydy}.  Thus, the projected quantity defined as the observed fluctuations of the galaxy number counts at a given sky position $\hat{n}$ is related to the three-dimensional matter density fluctuations $\delta_m(z, \chi\hat{n})$ as
\begin{equation}
\delta_g(\hat{n})=\int_0^{\chi_h} d\chi b(\chi)n(\chi)\delta_m(z(\chi),\chi\hat{n}),
    \label{eq:related_to_matter}
\end{equation}
where $b(\chi)$ is the galaxy bias and $n(\chi)$ the normalised distribution of galaxies. Then, the kernel in \autoref{eq:Harmonic_limber} takes the form
\begin{equation}
W^{\delta_g}(\ell,\chi)\equiv W^{\delta_g}(\chi)=n(\chi)b(\chi)\;.
\label{eq:GC_kernel}
\end{equation}

We do not consider any other correcting term on top of the galaxy density field, like magnification bias or redshift-space distortions which both are subdominant in our analysis and are relevant for tomographic analysis and narrow redshift bins, respectively \citep{Tanidis_mag}.

\subsection{CMB Lensing}
\label{subsec:CMBLens}

The convergence field $\kappa(\hat{n})$ is defined as the distortion of the CMB photon trajectories due to the gravitational potential caused by the underlying dark matter field \citep{LEWIS20061}. This is proportional to the divergence of the deflection in the photon arrival angle $\vec \alpha$ as: $\kappa\equiv-{\nabla} \cdot {\vec \alpha}/2$. Thus, $\kappa$ is an unbiased tracer of the matter density fluctuations $\delta_m(z, \chi\hat{n})$ and is related to them as
\begin{equation}
\kappa(\hat{n})=\int_0^{\chi_\star} d\chi \frac{3\om H_0^2}{2c^2}[1+z(\chi)]\chi \frac{\chi_\star-\chi}{\chi_\star}\delta_m(z(\chi), \chi\hat{n}),
    \label{eq:CMB_Lens}
\end{equation}
with $c$ the speed of light, $\om$ the matter fraction at present, $H_0$ the Hubble constant in units of $\mathrm{km\,s^{-1}\,Mpc^{-1}}$, and $\chi_\star$ the comoving distance at the last scattering surface corresponding to $z_\star\approx1100$. The radial kernel in this case takes the form
\begin{equation}
W^\kappa(\ell,\chi)=L(\ell) \frac{3\Omega_m H_0^2}{2c^2}[1+z(\chi)]\chi \frac{\chi_\star-\chi}{\chi_\star}
    \label{eq:CMB_Lens_kernel},
\end{equation}
where the factor $L(\ell)$ reads
\begin{equation}
    L(\ell)=\frac{\ell(\ell+1)}{(\ell+1/2)^2},
\end{equation}
which is only relevant (starts to deviate from unity) at $\ell\lesssim 10$. This term accounts for the fact that $\kappa$ is related to $\delta_m$ through the angular Laplacian of the {lensing} potential $\phi$ as: 
$\kappa(\hat{n})=-{\nabla^2}\phi(\hat{n})/2$.

\section{Data}
\label{sec:dataset} 

\subsection{EMU Pilot Survey 1}
\label{subsec:EMUPS}

The radio continuum galaxy sample used here is the Pilot Survey 1 of the Evolutionary Map of the Universe \citep[EMU PS1;][]{EMUPS1}. EMU will cover the complete southern sky within five years and will observe several tens of million sources \citep{norris2011, johnston2007, Johnston2008}. Here we use the first pilot data covering a contiguous patch of $\sim 270 \text{ deg}^2$, observed at 944 MHz, at a spatial resolution of 11-18 $\text{arcsec}$ and reaching a depth of 25-30 $\mu \text{Jy/beam rms}$. The resulting catalogue corresponds to roughly $\sim$ 200,000 sources for the full sample. The exact number slightly differs depending on the source finding algorithm used and it is further reduced after applying flux density cuts as we discuss in \autoref{subsubsec:Source finding algorithms}.

\subsubsection{Source finding algorithms and flux density cuts}
\label{subsubsec:Source finding algorithms}

The first source finding algorithm output we used is from the \texttt{Selavy} software \citep{Whiting_Humphreys_2012, Whiting_2017}. The tool identifies pixels that have emission  above a certain threshold, in this case 5 sigma (5 times the local rms in the image, \texttt{Selavy} variable \texttt{snrCut=5}) using the flood-fill technique, and groups the pixels that lie next to each other together into a single `island'. Then, if there are nearby pixels that lie above a lower threshold (in our case 3 times the rms, \texttt{growthThreshold=3}), the island can be ‘grown’ to encompass these pixels also. As discussed in \citet{Whiting_Humphreys_2012}, this `growing’ can lead to nearby sources being merged with the source under-consideration. Finally then, it fits Gaussian components to peaks of emission within the islands (\texttt{Fitter.doFit=true}). We note that we use the estimate of the total flux density of each source by summing over the Gaussians that have been fit to the components, for a more accurate integrated flux density estimate. 

\begin{figure}
\centering
\includegraphics[width=0.5\textwidth]{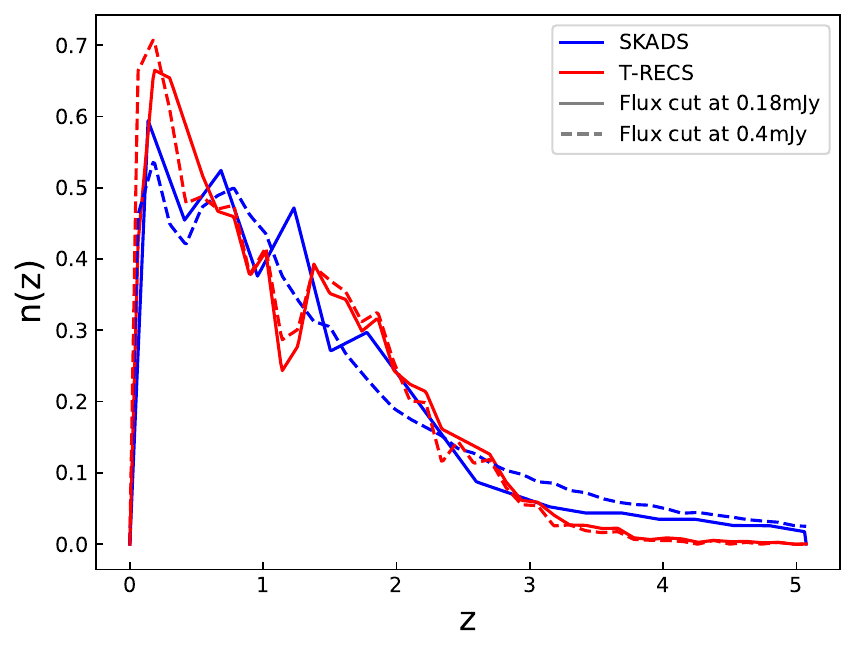}
\caption{The normalised redshift distributions of radio continuum galaxies as estimated from the simulations \texttt{SKADS} (blue) and \texttt{T-RECS} (red) at the flux density cuts 0.18mJy (solid) and 0.4mJy (dashed).}
\label{fig:redshift_dists}
\end{figure}

\begin{figure*}[hbt!]
\centering
\includegraphics[width=0.5\textwidth]{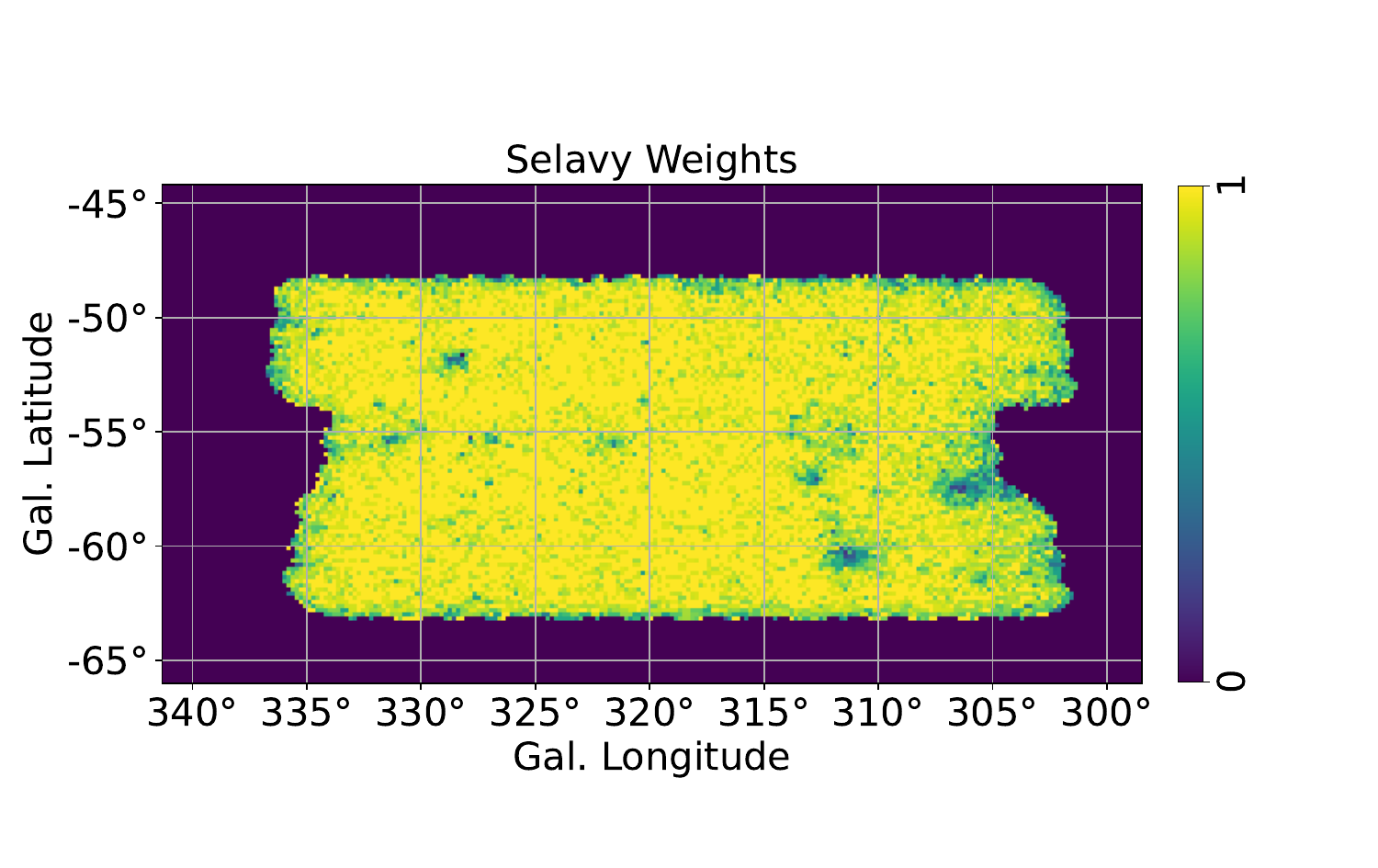}\includegraphics[width=0.5\textwidth]{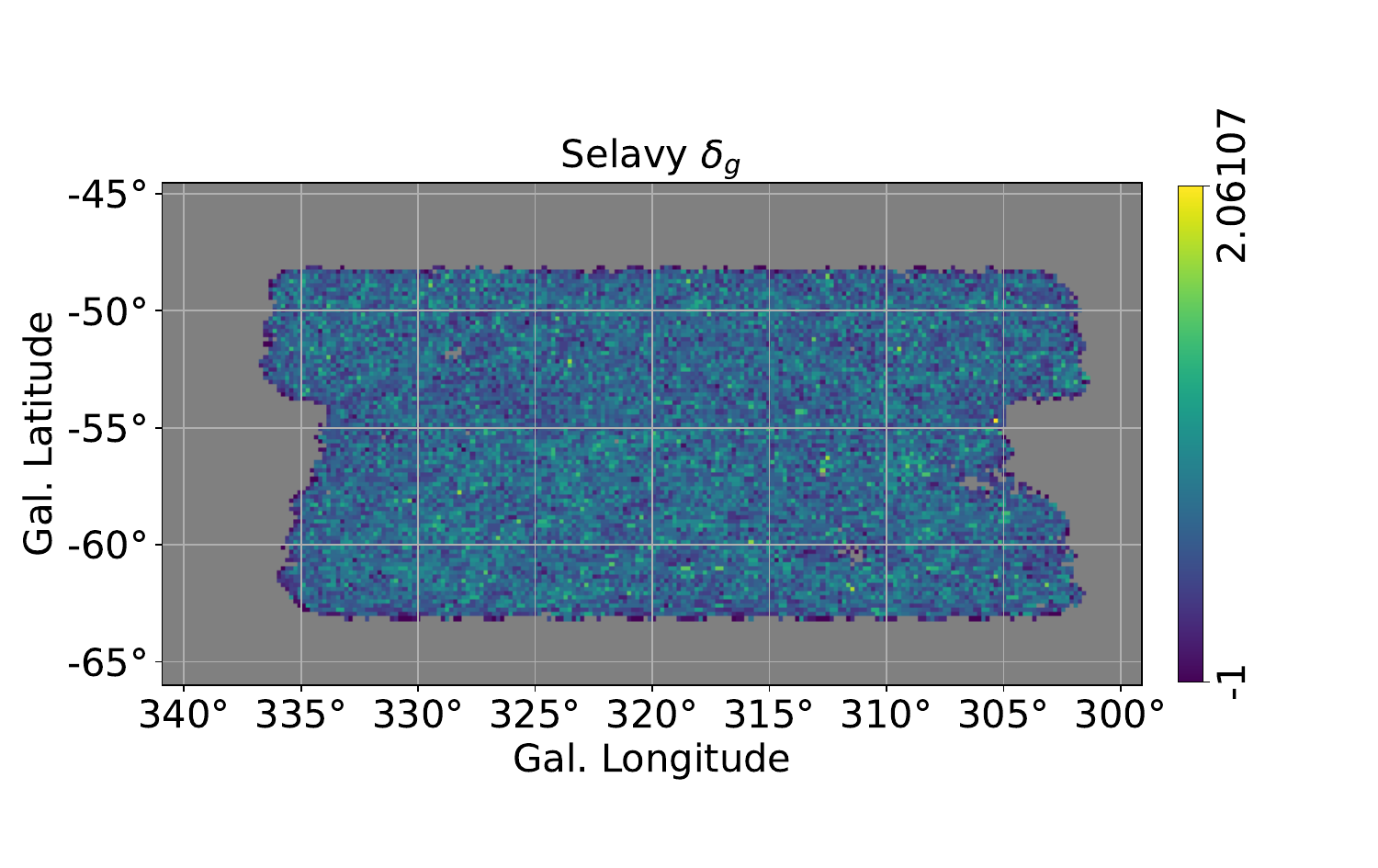}\\
\includegraphics[width=0.5\textwidth]{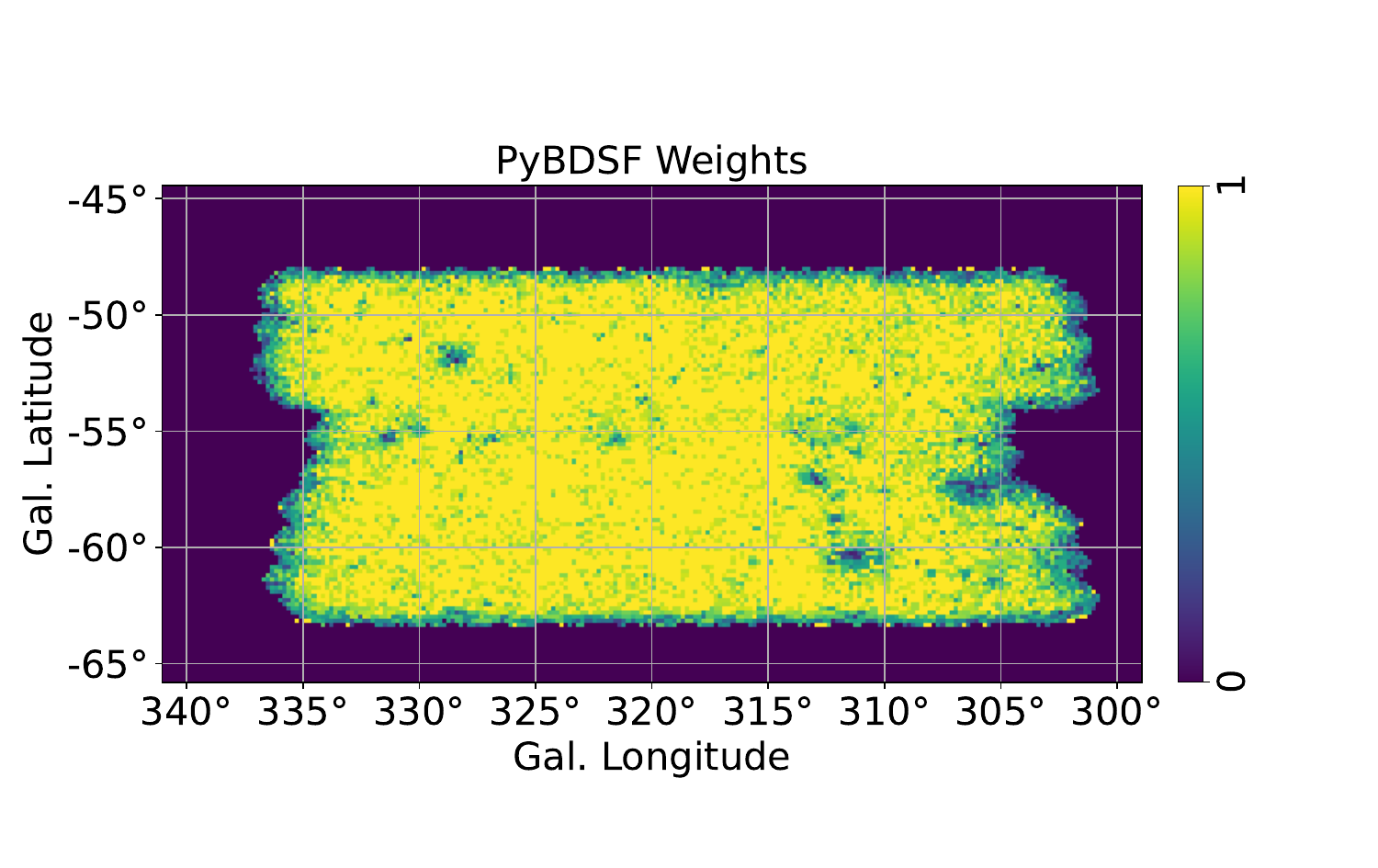}\includegraphics[width=0.5\textwidth]{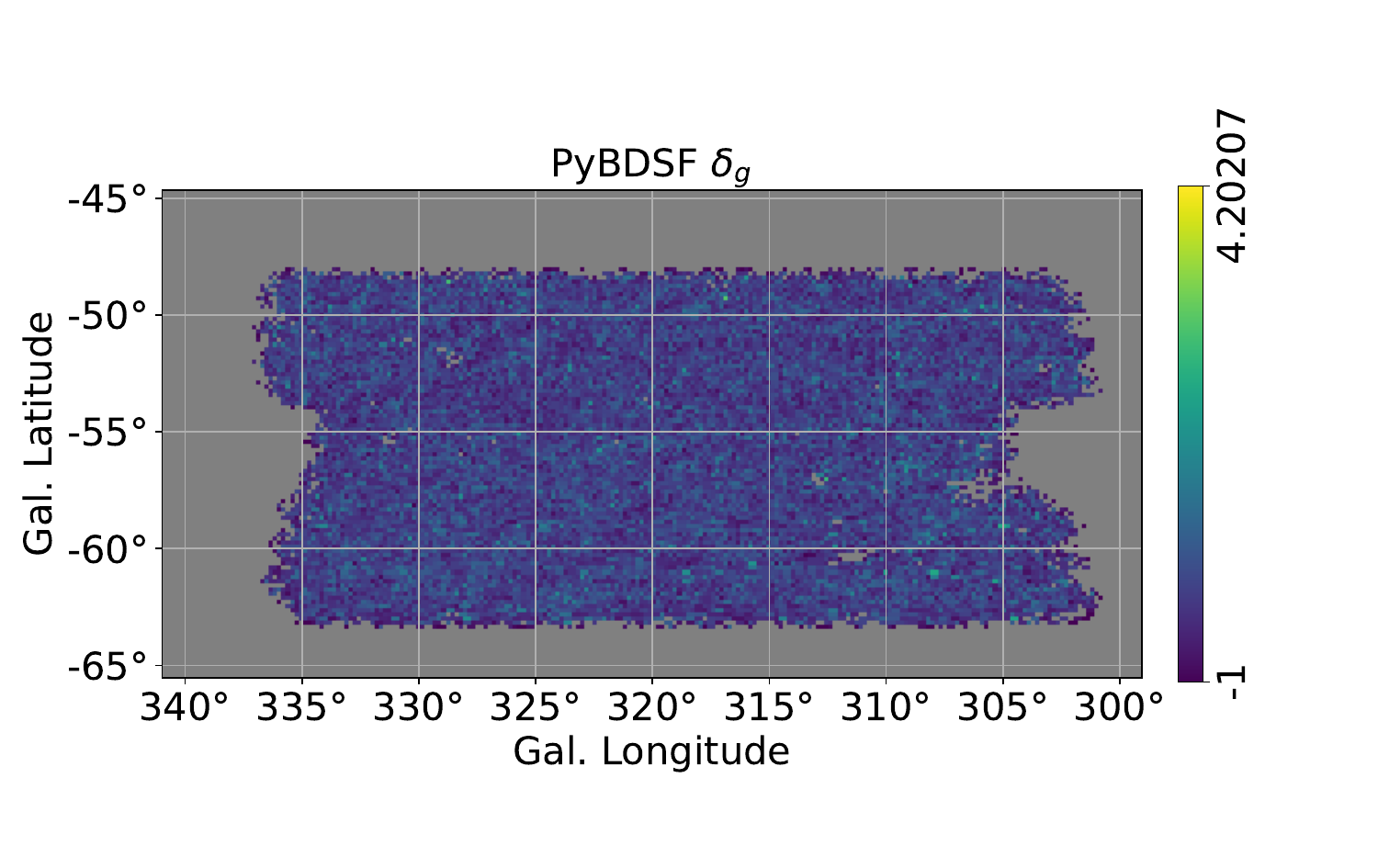}\\
\includegraphics[width=0.5\textwidth]{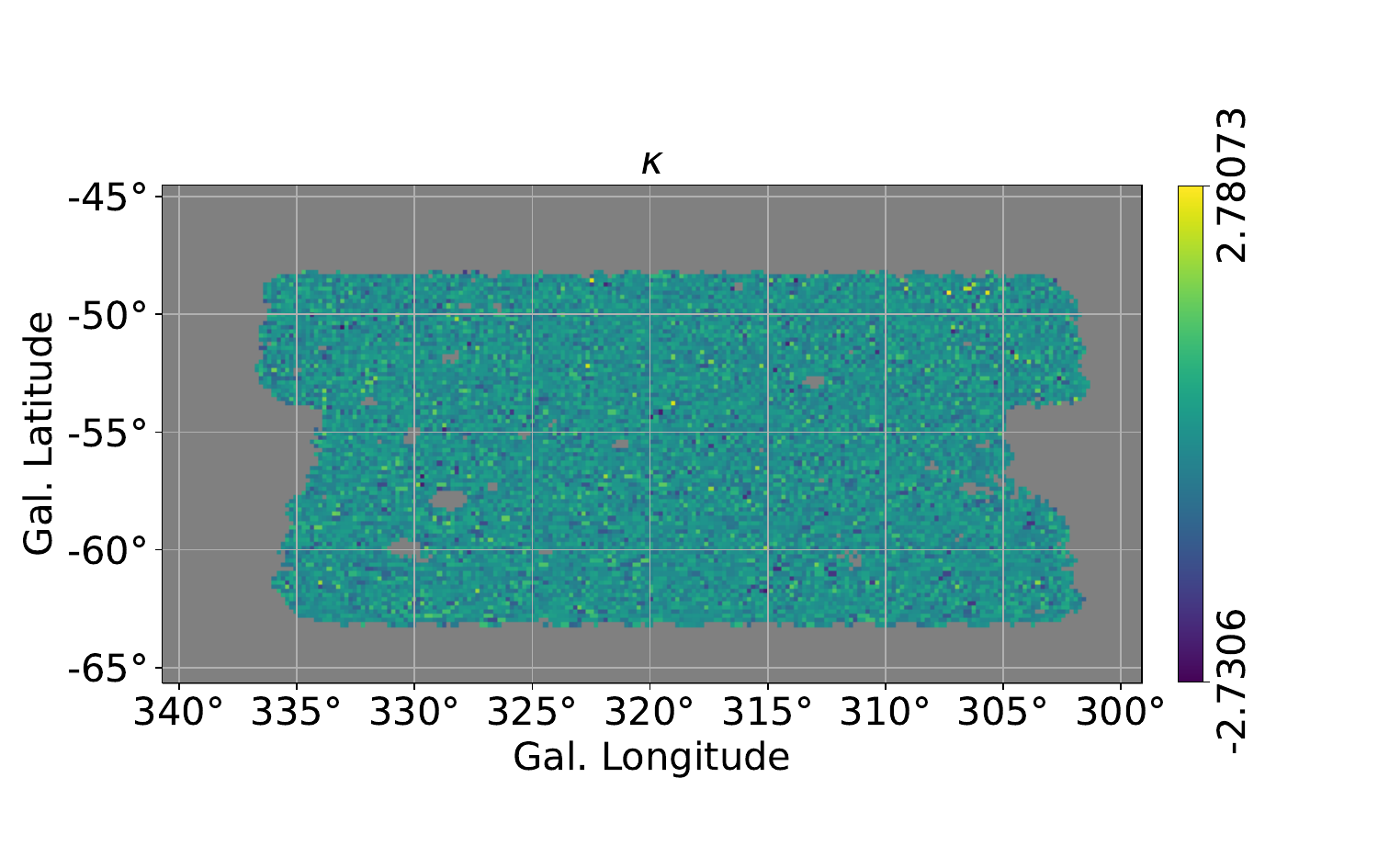}
\caption{A list of maps that was used in our work. \textit{Top left:} The weights mask for \texttt{Selavy}. \textit{Top right:} The galaxy overdensity map for \texttt{Selavy}. \textit{Middle left:} The weights mask for \texttt{PyBDSF}. \textit{Middle right:} The galaxy overdensity map for \texttt{PyBDSF}.  \textit{Bottom:} The CMB convergence map. All the galaxy maps here are for the flux density cut at $0.18$ mJy, while for the cut at $0.4$ mJy, they look similar. In the overdensities and convergence panels, the mask is shown with grey color.}
\label{fig:footprints}
\end{figure*}

At this point we note that we consider only the \texttt{Selavy} island sample and not the \texttt{Selavy} component sample for the cosmological analysis described in \autoref{sec:results}. We make this choice due to the fact that as sources can be quite extended in the images, different components can be generated by the same radio galaxy. This can affect the clustering statistics at small scales $<0.1^\circ$ (see again \citealt{EMUPS1}). Even though we use the islands catalogue, there still may exist residual biases we need to account for (see discussion in \autoref{subsec:like}). We also cross-check that the clustering measurements we discuss in \autoref{subsec:detectionsig} using the \texttt{Selavy} island catalogue are in good agreement with the machine-learning based morphological classification of EMU-PS radio catalogue compiled with the \texttt{Gal-DINO} pipeline \citep{gupta2024rgcatdetectionpipelinecatalogue}.

The other source finder algorithm we used to generate the catalogue is \texttt{PyBDSF}\footnote{https://pybdsf.readthedocs.io/en/latest/index.html} \citep{2015ascl.soft02007M}. To do this, we set a threshold which determines which pixels contribute to an island of emission (\texttt{thresh\_isl}) to be 3$\sigma$ and the threshold for source detection (\texttt{thresh\_pix}) to 5$\sigma$. We additionally include a specification that the background mean level should be zero (\texttt{mean\_map=`zero'}) and specify the box size and step size used to generate the rms map (\texttt{rms\_box = (150,30)}). From running \texttt{PyBDSF} over the image we record the rms map, to generate random sources, alongside the output source and Gaussian catalogues.

In addition, we consider only the galaxies with flux density density brighter than 0.18 mJy. The choice is based on the fact that for sources brighter than this value, the source counts in the previous models and simulations \citep{Mancuso2017, Bonaldi2019} are in agreement with the EMU PS1 island catalogue \citep{EMUPS1}. In order to test the robustness of our cosmological analysis on the galaxy sample, we also consider a more rigorous flux density cut at $0.4$ mJy. We perform these cuts both on \texttt{Selavy} and \texttt{PyBDSF} catalogues. The number of galaxies after the flux density cuts and the maps are discussed in \autoref{subsec:pseudo-cl}.

\subsubsection{Redshift distributions}
\label{subsubsec:red_dist}
As we can appreciate from \autoref{eq:GC_kernel}, to estimate the kernel of the radio continuum galaxy sample we need to obtain an accurate model for the galaxy number distribution as a function of redshift. To achieve that, we make use of two of the largest extragalactic radio galaxy simulations; the European SKA Design Study (hereafter \texttt{SKADS}) Simulated Skies \citep{Wilman2008} and the Tiered Radio Extragalactic Continuum Simulation \citep[hereafter \texttt{T-RECS};][]{Bonaldi2019}. Also, we consider for both simulations AGNs and SFGs contributions, which constitute the main tracers of the galaxy populations in the radio surveys. In \autoref{fig:redshift_dists}, we show the redshift distributions for \texttt{SKADS} and \texttt{T-RECS} and for the flux density cuts at $0.18$ and $0.4$ mJy. The distributions are not affected considerably by the flux density cuts and both \texttt{SKADS} and \texttt{T-RECS} are peaked at $z\sim0.5$, after which they fall slowly up to high redshifts. Nonetheless, we can appreciate that the \texttt{SKADS} has longer tail at high redshift, while the \texttt{T-RECS} is more localized at $z\sim0.5$. This can affect power spectra fits to the data, since samples with broader redshift distributions wash out their structure information, decreasing in this way the power amplitude and in turn increasing the galaxy bias $b_g$. However, as we discuss in \autoref{subsec:Gal_bias_results}, this difference does not affect the results significantly {(shift of $\sim0.2\sigma$)}. Thus, for our baseline results of \autoref{subsec:Gal_bias_results} and \autoref{subsec:sigma8}, we use the \texttt{SKADS} distribution. 

{It is important to stress at this point that \texttt{SKADS} and \texttt{T-RECS} have both similarities (all of them considering AGNs and SFGs) and differences (empirical models for the former and more detailed population models for the latter) and therefore, it should not be surprising that the redshift distributions look similar. The comparison we make between them in this work (\autoref{app:trecs_analysis}) certainly should not be seen as a robustness systematic test but rather as an indicative comparison between the state-of-the art radio continuum simulation codes given the large uncertainty in the redshift distribution of the radio galaxies. In fact, there is a series of ongoing parallel works aimed to constrain both the peak and the tail of the redshift distribution of the EMU radio sample by using cross-correlation with the Dark Energy Survey \citep[DES,][]{Abbott_2016} optical galaxies \citep{Saraf_2025} and the Euclid telescope \citep{euclidcollaboration2024euclidiovervieweuclid} deep fields \citep{Kalus_2025}. Also, a cross-matching study that will further help in the modelling of the redshift distribution is planned in the future.} 

\subsection{\textit{Planck} PR4}
\label{subsec:planck_pr4}

We use the publicly available CMB lensing convergence map $\kappa$ from the \textit{Planck} PR4 data \citep{Carron_2022}. This map is constructed using an improved lensing quadratic estimator and contains $\sim$8\% more data than the previous release of 2018 \textit{Planck} PR3 \citep{2020A&A...641A...8P}. The harmonic coefficients of the $\kappa$ mean-field subtracted map are transformed to a \texttt{HEALPix} map with $N_\text{side}=512$ corresponding to a pixel size of $\sim$6.9 arcmin. This resolution is also used for the galaxy overdensity maps which is discussed in \autoref{subsec:pseudo-cl} and it is considered to be accurate enough for the scales we probe in this work. The convergence map covers $\sim$67\% of the sky and fully overlaps with the footprint of the EMU PS1 map. The convergence map has a few holes that remove less than 1\% of the EMU-PS1 footprint (see \autoref{fig:footprints}).



\section{Methodology}
\label{sec:methodology}
\subsection{Pseudo-$C_\ell$s}
\label{subsec:pseudo-cl}

The harmonic-space coefficients and the spectrum of \autoref{eq:Harmonic_definition} are defined under the full-sky assumption. In reality, we are able to observe only a part of the sky. This is true both for the radio continuum galaxy maps and the CMB lensing convergence map as we have seen in \autoref{subsec:EMUPS} and \autoref{subsec:planck_pr4}. Thus, the measured values of harmonic coefficients differ from the full-sky ones leading to the pseudo-$C_\ell$ spectrum which accounts for the partial sky. We do this by using the python package \texttt{NaMaster} \citep{Alonso19}. 


{To construct the weight maps for EMU-PS1, we create a mask that accounts for the rms of the EMU PS1 mosaic. To do so, we create a galaxy random catalogue by following the method described in \cite{Hale18}. We start by drawing uniform random angular positions, and random flux densities from the SKADS simulation \citep{Wilman2008} at a frequency of 1.4GHz, scaled to 944 MHz. For each of the catalogs (Selavy or PyBDF), an rms image is produced respectively. These RMS images allow us to include the observational noise in each position of the map. We then only select random galaxies with flux densities with a significance $5\sigma$ above the rms level, given by each catalogue rms map, of the corresponding angular position. Once we have a random catalogue, we apply a flux density cut for the corresponding galaxy sample. The weights mask is just the ratio between the number of random galaxies in a given \texttt{HEALPix} pixel and the number of random galaxies from the original uniform randoms (before the rms flux density cut). The randoms are created in a set of realizations, producing $20000$ uniform randoms for each realisation. The final number of random galaxies used was selected by checking the stability of the spectra measurements for a given number of realizations. We found that the pseudo-$C_\ell$s spectra are robust when the number of realisations to produce the randoms is above $500$.}

The definition of the observed fluctuations in the galaxy number counts now becomes,  
\begin{equation}
\delta_g(\hat{n})=\frac{N_g(\hat{n})}{\Bar{N}_g w_g(\hat{n})}-1,
    \label{eq:overdensity_with_weights}
\end{equation}
with $N_g(\hat{n})$ the number of galaxies in the pixel position $\hat{n}$ and $w_g(\hat{n})$ the weights in the same pixel position. $\Bar{N}_g$ is the weighted mean number of sources per pixel in our samples and reads: $\Bar{N}_g=\left\langle  N_g(\hat{n})\right\rangle_n/{\left\langle  w_g(\hat{n})\right\rangle}_n$, with $\left\langle . \right\rangle_n$ denoting the mean over all the pixels in the map. We also avoid heavily masked pixels by setting $w_g(\hat{n})$ and $\delta_g(\hat{n})$ pixels to zero where $w_g(\hat{n})<0.5$. The final number of galaxies for the \texttt{Selavy} catalogue is 166,801 at $>0.18$ mJy and 83,222 at $>0.4$ mJy, while for \texttt{PyBDSF} are 188,034 and 89,320, respectively. The weight footprints and galaxy overdensity maps for \texttt{Selavy} and \texttt{PyBDSF} are shown in  \autoref{fig:footprints} with the latter having 2\% larger footprint than the former. This is due to the fact that \texttt{Selavy} has a stricter limit on the acceptable weights at the edges of the footprint truncating slightly the coverage.

After the construction of the overdensity maps we transform the weight maps to binary masks {and we couple the spectra with them using \texttt{NaMaster}. We verify that our results are stable when we perform this.} The pseudo-$C_\ell$ harmonic-space spectrum \citep{Hivon_2002} is defined as,
\begin{equation}
\bar{C}^{XY}_{\ell,s+n}=\frac{1}{2\ell+1}\sum_{m=-\ell}^\ell \bar{X}_{\ell m} \bar{Y}^*_{\ell m},
    \label{eq:pseudoCl}
\end{equation}
where $\bar{X}_{\ell m}$ and $\bar{Y}_{\ell m}$ denote the partial sky harmonic coefficients  of the fields receiving contributions from signal and noise $s+n$. The observed harmonic-space spectrum is the ensemble average $\tilde{C}^{XY}_{\ell,s+n}=\langle \bar{C}^{XY}_{\ell,s+n} \rangle$ and is related to the true signal $S_\ell$ (see again \autoref{eq:Harmonic_definition}) via 
\begin{equation}
\tilde{C}^{XY}_{\ell,s+n}=\sum_{\ell'} M^{XY}_{\ell\ell'} S^{XY}_{\ell'}+\tilde{N}^{XY}_\ell,
    \label{eq:mixing_spectra}
\end{equation}
where $\tilde{N}^{XY}_\ell=\delta^{\rm K}_{XY}\Omega_\text{p}\bar{w}_g/\bar{N}_X$ is the shot noise \citep{Nicola_2020} with $\bar{w}_g$ the average value of the mask across the sky (see \autoref{eq:overdensity_with_weights}) and $\Omega_\text{p}$ the pixel area in units of steradians. The noise needs to be subtracted for auto-correlations to obtain the masked signal $\tilde{C}^{XY}_{\ell, s}=\tilde{C}^{XY}_{\ell, s+n}-\tilde{N}^{XY}_\ell$. By rescaling it with the survey sky fraction $f_\text{sky}$, we can get an estimate of the true spectrum as $\tilde{C}^{XY}_{\ell}\equiv S^{XY}_\ell=\tilde{C}^{XY}_{\ell, s}/f_\text{sky}$, which is a good approximation for fairly flat power spectra, as we consider here \citep{Nicola_2021}.  The quantity $M_{\ell\ell'}$ is the mode coupling matrix \citep{Peebles1973} due to the masked area and it is defined as,
\begin{equation}
M^{XY}_{\ell\ell'} = 
\frac{2 \ell^{\prime} + 1}{4 \pi}
\sum_{\ell^{\prime \prime}} (2 \ell^{\prime\prime} + 1)
{W_{\ell^{\prime \prime}}^{XY}  
\left (
\begin{array}{lll}
\displaystyle{\ell} & \displaystyle{\ell^{\prime}} & \displaystyle{\ell^{\prime \prime}} \\
& & \\
\displaystyle{0} & \displaystyle{0} & \displaystyle{0} \\
\end{array}
\right )^2} \ ,
    \label{eq:mixing_matrix}
\end{equation}
with $W_\ell^{XY}$ the spectra of the masks which read,
\begin{equation}
W_\ell^{XY} = \frac{1}{2\ell+1}\,\sum_{m=-\ell}^{\ell} w_{\ell m}^X {w_{\ell m}^{Y\ast}}, 
    \label{eq:weight_spectra}
\end{equation}
where $w_{\ell m}^X$ and $w_{\ell m}^Y$ are the spherical harmonic coefficients of the masks of the fields under study. 

As we elaborate in \autoref{subsec:like} we need to compare the measured spectra with the model spectra from theory that we saw in \autoref{eq:Harmonic_limber}. To do this, we also account for the partial sky effect in the theory spectra by applying the same coupling matrix convolution and the rescaling correction as,
\begin{equation}
\tilde{S}^{XY}_\ell=\left(\sum_{\ell'} M^{XY}_{\ell\ell'} S^{XY}_{\ell', \text{th}}\right)/f_\text{sky}.
    \label{eq:mixing_theory}
\end{equation}

\subsection{Matter Spectrum and Galaxy Bias Models}
\label{subsec:Matter_PS_Galaxy_Bias_Models}

We consider both a linear and a non-linear matter power spectrum $P_{mm}$ for the theory harmonic-space spectrum of \autoref{eq:Harmonic_limber}. To obtain the linear model we use the Boltzmann solver \texttt{CAMB} \citep{Lewis2000} and we get the non-linear model from it with \texttt{HALOFIT} \citep{Smith2003,Takahashi2012}. Unless otherwise stated, we use the fiducial cosmology best-fit values by \cite{Planck2020}, which are: present day cold dark mater fraction, $\oc=0.26503$; present day baryon fraction, $\ob=0.04939$; rms variance of linear matter fluctuations at present in spheres of $8\,h^{-1}\,\mathrm{Mpc}$, $\sigma_8=0.8111$; dimensionless Hubble constant $h\equiv H_0/(100\,\mathrm{km\,s^{-1}\,Mpc^{-1}}) = 0.6732$ and primordial power spectrum spectral index, $n_s=0.96605$. The theoretical calculations in this work are done using the code \texttt{CosmoSIS} \citep{Zuntz2015}.

Regarding the galaxy bias redshift evolution we consider two models \citep{Alonso2021}:
\begin{itemize}
    \item A constant galaxy bias model with $b(z)=b_g$ which represents a simple scenario, where the growth evolution with time of the galaxy clustering follows that of the matter fluctuations.
    \item A constant amplitude galaxy bias model with $b(z)=b_{g}/D(z)$ which evolves with the inverse of the linear growth factor defined as: $D(z)=[P^\text{lin}_{mm}(k,z)/P^\text{lin}_{mm}(k,0)]^{1/2}$ in the linear regime $k\rightarrow{0}$. This model, though still simple with one parameter as well, preserves its large-scale properties unchanged and remains fixed at early times (since at linear scales $\delta_m\propto D$). At the same time, it is able to reproduce the expected rise in $b(z)$ at high redshift for a flux density-limited galaxy sample \citep[{e.g.}][]{Bardeen:1985tr,Mo1996,Tegmark_1998, Coil2004}.
\end{itemize}

On top of these models, we also test another more flexible case, that of a quadratic galaxy bias model: $b(z)=b_0+b_1z + b_2z^2$ with three parameters $\{b_0, b_1, b_2\}$. However, as we later discuss in \autoref{app:quadratic_model} the results of this model are consistent with those of the constant galaxy bias and, therefore we do not use it in our fiducial analysis of \autoref{sec:results}.

Finally, in our pipeline we consider scales up to $\ell_\text{max}=500$. This corresponds to $k_\text{max}\sim 0.15$ Mpc$^{-1}$ at $z_\text{med}\sim 1$ which is the rough median redshift for both distributions (in particular,  $z_\text{med}\sim0.98$ for \texttt{T-RECS} and $z_\text{med}\sim 1.1$ for \texttt{SKADS}). In this mildly non-linear regime, the linear galaxy bias model is a good approximation, while we can neglect non-Gaussian contributions to the covariance matrix {\citep[e.g.][]{Smith_2007, Cooray_2004}}.

\subsection{Covariance Matrix and Likelihood}
\label{subsec:like}

Assuming that $\kappa$ and $g$ are random variables, we can write the analytical covariance matrix $\mathbf K$ terms for the auto-correlation $gg$ and the cross-correlation $g\kappa$ spectra as follows,
\begin{equation}
\mathbf K=\left [
\begin{array}{ll}
\displaystyle{\mathbf K^{gg,gg}} & \displaystyle{\mathbf K^{gg,g\kappa}} \\

\displaystyle{(\mathbf K^{gg,g\kappa})^{\sf T}} & \displaystyle{\mathbf K^{g\kappa,g\kappa}} \\
\end{array}
\right ],
    \label{eq:covs}
\end{equation}
with each sub-block taking the form,
\begin{multline}
\mathbf K^{gX,gY}_{\ell\ell'}=\frac{\delta_{\ell\ell'}}{(2\ell+1)\Delta\ell f_\text{sky}^{gX,gY}}[(\tilde{C}^{gg}_\ell+N^{gg}_\ell)(\tilde{C}^{XY}_{\ell}+N^{XY}_{\ell})\\
+(\tilde{C}^{gX}_\ell+N^{gX}_\ell)(\tilde{C}^{gY}_{\ell}+N^{gY}_{\ell})],
    \label{eq:cov_blocks}
\end{multline}
where $X$ and $Y$ can both be $g$ or $\kappa$ and $\Delta\ell$ the multipole binwidth. The sky fractions read: $f_\text{sky}^{gX,gY}=\sqrt{f_\text{sky}^{gX}\cdot f_\text{sky}^{gY}}$ and $f_\text{sky}^{gg}\approx f_\text{sky}^{g\kappa}$. We also bin the measured {masked and rescaled} $\tilde{C}^{XY}_\ell$ as well as the theory $\tilde{S}^{XY}_\ell$ power spectra with $N_\ell$=11 multipoles, linearly from\footnote{The validity of the Limber approximation at $\ell_\text{min}$=2 for a single redshift bin of an EMU-like survey has been confirmed by \cite{Tanidis_mag} and \cite{Bahr-Kalus2022}} $\ell_\text{min}=2$ to $\ell_\text{max}=500$. In addition, we verify that our results using the analytical covariance in \autoref{sec:results} are robust by comparing them with the numerical covariance which is described in \autoref{app:mocks_cov}.


\begin{figure*}[hbt!]
\centering
\includegraphics[width=0.5\textwidth]{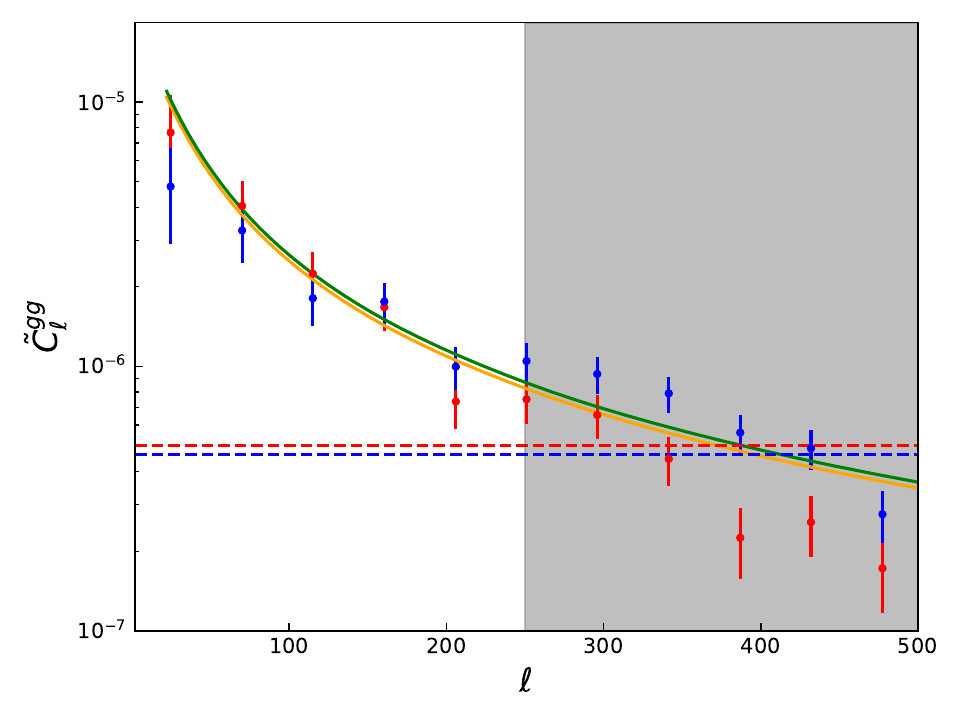}\includegraphics[width=0.5\textwidth]{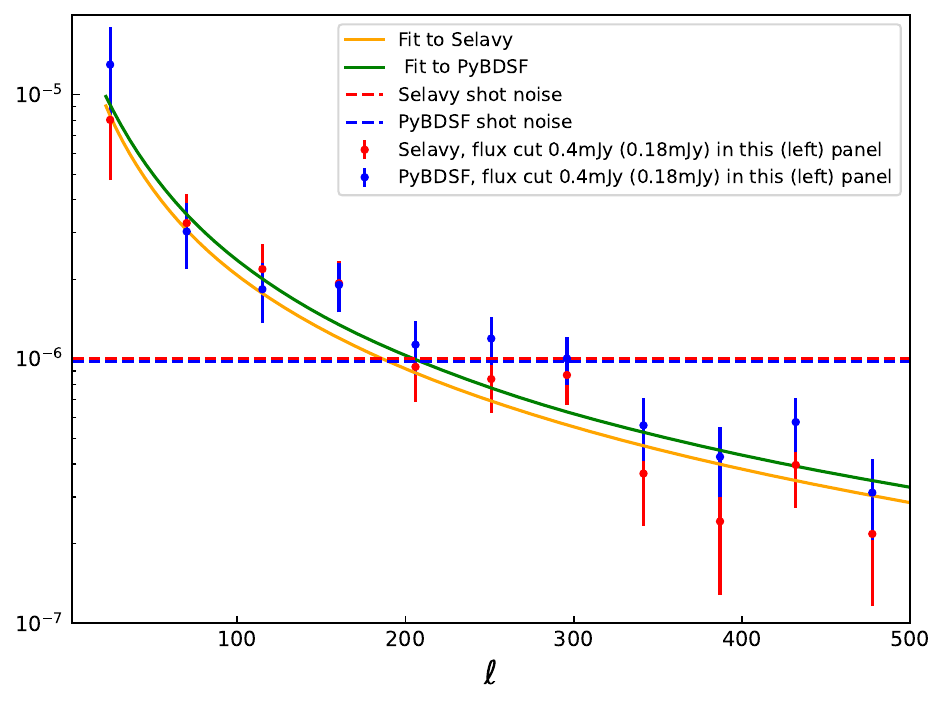}\\
\includegraphics[width=0.5\textwidth]{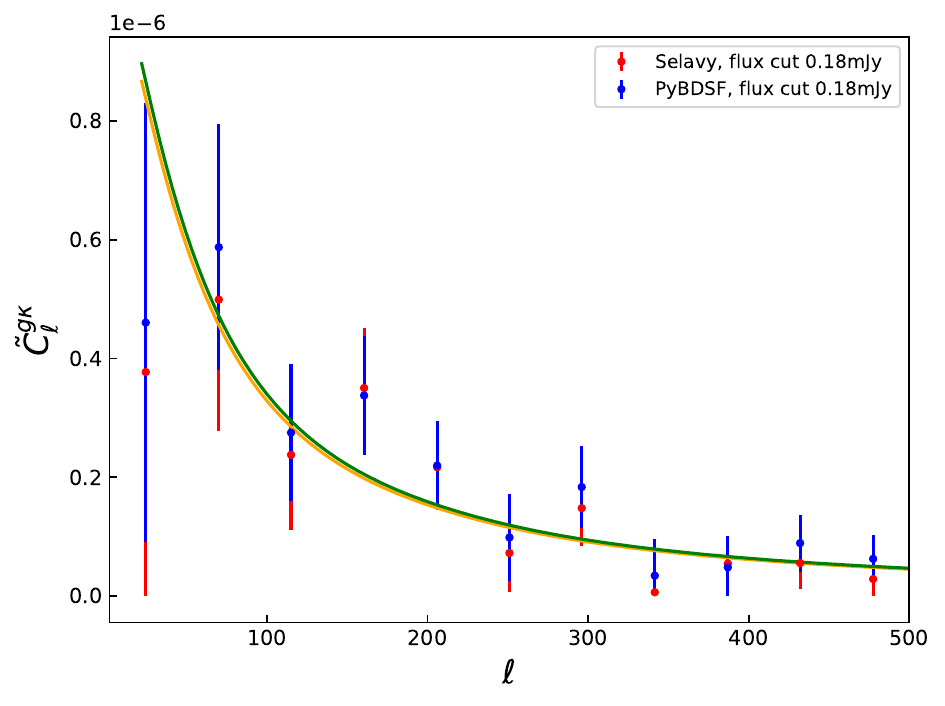}
\caption{The auto-correlation $\tilde{C}^{gg}_\ell$ for the flux density cut at 0.18mJy (top left panel) and 0.4mJy (top right panel). Red and blue points along with their 1$\sigma$ uncertainties, correspond to the \texttt{Selavy} and \texttt{PyBDSF} catalogues. Their corresponding fitted theory models are denoted with orange and green curves, respectively, which are estimated assuming the \textit{Planck} best-fit values \citep{Planck2020}, the \texttt{SKADS} redshift distribution and \texttt{HALOFIT} power spectrum. The colourful horizonal dashed lines are shot noise estimates for the two catalogues, and the grey shaded area (top left panel) denotes the scale cut at $\ell=250$ for the flux density cut at 0.18mJy. The bottom panel shows the cross-correlation of galaxies with the CMB lensing convergence $\tilde{C}^{g\kappa}_\ell$ at the flux density cut 0.18mJy.}
\label{fig:scale_cut_flux_0p18}
\end{figure*}


Assuming that the spectra follow a Gaussian {distribution}, we can use the log-likelihood,
\begin{equation}
\chi^2(\mathbf{q})=\sum_{\ell,\ell'}[\bm{d}_\ell-\bm{t}_\ell(\bm{q})]^{\sf T}\,\mathbf{K}_{\ell\ell'}^{-1}\,[\bm{d}_{\ell'}-\bm{t}_{\ell'}(\bm{q})], 
    \label{eq:loglike}
\end{equation}
where $\bm{d}_\ell=\{\tilde{C}^{gg}_\ell, \tilde{C}^{g\kappa}_\ell\}$ and $\bm{t}_\ell=\{\tilde{S}^{gg}_\ell, \tilde{S}^{g\kappa}_\ell\}$ denote the data and theory model vectors and $\bm{q}$ the parameter set of interest we want to fit. In our analysis we aim to constrain the galaxy bias $b_g$ and $\sigma_8$. For both parameters we assume flat priors $b_g\in(0.01, 10)$ and $\sigma_8\in(0.01, 1.6)$. To estimate the posterior distributions of the parameters we use publicly available Bayesian-based sampler \texttt{emcee} \citep{ForemanMackey2013}.

\section{Results}
\label{sec:results}

\subsection{Differences between the source finding algorithms and deviations from shot noise}

As discussed in \autoref{subsubsec:Source finding algorithms} radio surveys can have multi-component structures that could affect the power spectrum and the Poissonian shot noise. At this point, we discuss the main difference between the two source finding algorithms, namely, the \texttt{Selavy} and \texttt{PyBDSF}, which we introduced in \autoref{subsubsec:Source finding algorithms}. In the island catalogue of \texttt{Selavy}, the algorithm categorizes as single objects, structures that are quite large. However, these large objects could, in fact, contain smaller sub-structures which could be part of the same extended object (multi-component object) or could belong to different sources. The \texttt{PyBDSF} algorithm is able to find these structures and categorize them as different sources. This can result, of course, in a larger power spectrum (more clustering) as measured by \texttt{PyBDSF} at small scales, where many smaller sources could correspond to a single large source for \texttt{Selavy}. Indeed, this is what we find for the measurements from the two catalogues in \autoref{subsec:detectionsig} Finding more clustering with the \texttt{PyBDSF} is not necessarily the correct thing, as the algorithm can incorrectly consider small sub-structures that may belong to a single galaxy, as different galaxies. Furthermore, there are additional effects like halo exclusion, and non-local and stochastic effects in galaxy formation \citep{Blake_2004, Tiwari2022}. 

All of these contributions can also induce deviations from Poissonian shot noise. To account for these contributions, we marginalise over an extra free amplitude parameter for the shot noise $A_\text{sn}$ \citep{Nakoneczny24} when we subtract it in the data auto-correlation $gg$ as $\tilde{C}^{XY}_{\ell, s}=\tilde{C}^{XY}_{\ell, s+n}-A_\text{sn}\tilde{N}^{XY}_\ell$, making the galaxy clustering auto-correlation sensitive to non-flat contributions. Based on the $\sim$20\% difference that was found between the island and component number of sources by \citealt{EMUPS1} and as we consider other potential biases as described above, we deem reasonable to consider an informative prior in the range $A_\text{sn}\in(0.8, 1.2)$, while we keep it fixed in the analytical covariance in \autoref{eq:covs}.

\subsection{Measurements and detection significance}
\label{subsec:detectionsig}



In the top left panel of \autoref{fig:scale_cut_flux_0p18}, we show the measured signal for the auto-correlation spectra $gg$ from the \texttt{Selavy} and \texttt{PyBDSF} catalogues for the flux density cut at $0.18$ mJy. With red points we denote \texttt{Selavy} and with blue points \texttt{PyBDSF} data, while the errorbars correspond to 1$\sigma$ uncertainties from the analytical covariance in \autoref{eq:cov_blocks}. The two catalogues are in agreement {(within 1$\sigma$)} until the scale $\ell\sim 250$, after which they start to deviate from each other. At $\ell \gtrsim 250$, the \texttt{PyBDSF} data have more power than \texttt{Selavy}. This can be attributed to the existence of multi-structures at small scales which are considered to be different objects by \texttt{PyBDSF}, and if they are close enough, as a single larger object by \texttt{Selavy}, as already explained in \autoref{subsec:like}. Therefore, we choose to apply a scale cut at $\ell \leq 250$, where the measurements from the two catalogues agree within $1\sigma$, and neglect smaller scales, in which the two algorithms {start to deviate} and the disentangling between the multi-sources and multi-components is really hard. Then we use a theory model using the \texttt{SKADS} redshift distribution and the \texttt{HALOFIT} non-linear matter power spectrum leaving free the $b_g$ and $A_\text{sn}$ parameters and fixing $\sigma_8$ in order to fit the \texttt{Selavy} and \texttt{PyBDSF} $gg$ spectra alone (the theory fits are with orange and green curves, respectively). It turns out the models fitting the two catalogues agree very well with each other {(with both models' best-fit values differ at $<0.1\sigma$ within their posteriors)}. 

In the bottom panel of \autoref{fig:scale_cut_flux_0p18} we see the $g\kappa$ cross-spectra between the radio galaxies and the CMB convergence $\kappa$ again for the two catalogues in a separate fit with $g\kappa$ data alone. It is evident that the data agree at all scales now up to $\ell=500$ {(well within $1\sigma$)} and the theory models agree as well {(again with both models' best-fit values differ at $<0.1\sigma$ within their posteriors)}. 

\begin{figure*}[hbt!]
\centering
\includegraphics[width=0.5\textwidth]{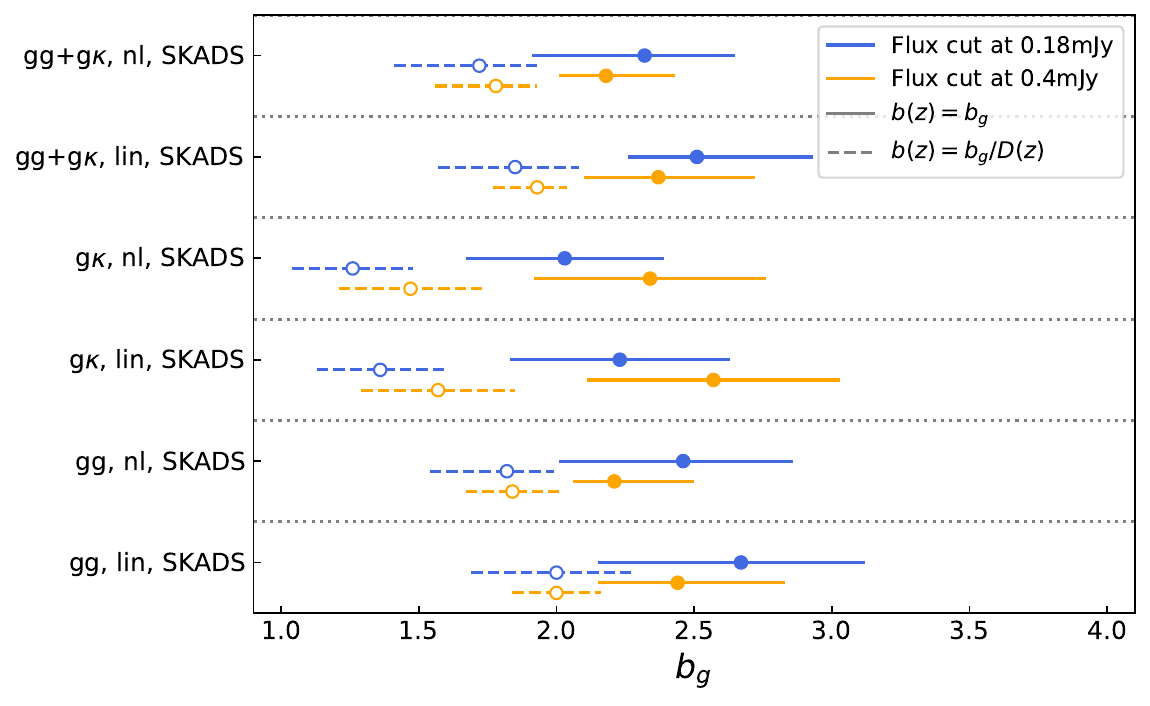}\includegraphics[width=0.45\textwidth]{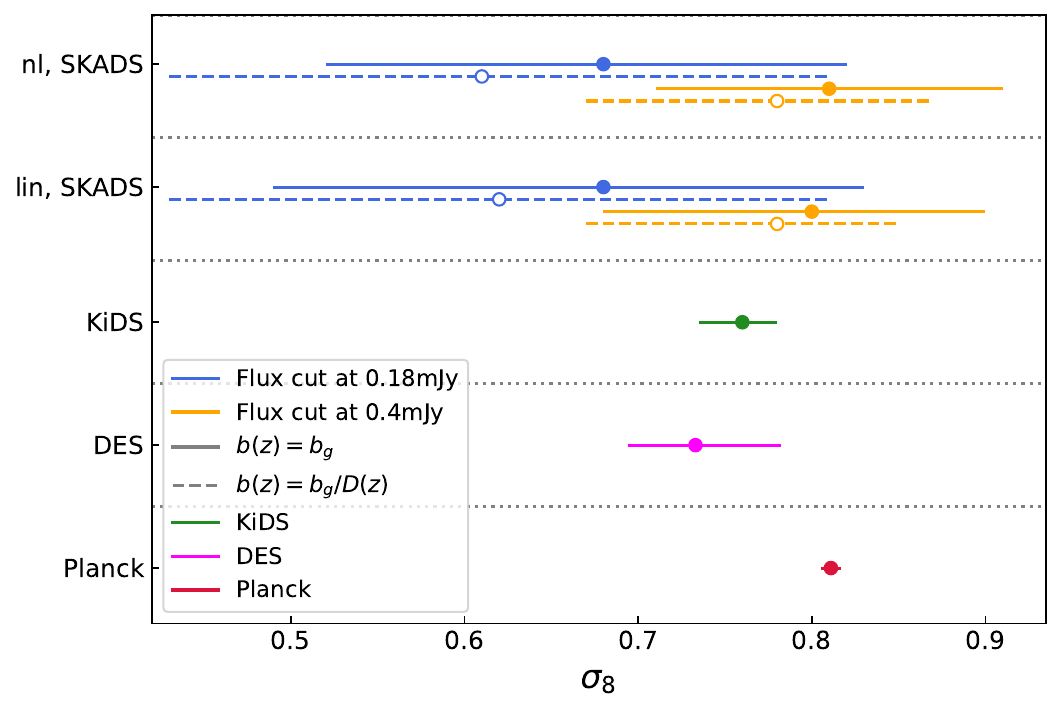}
\caption{\textit{Left:} The best-fit values along with their 68\% confidence intervals on the galaxy bias parameter $b_g$ for the auto-correlation $\tilde{C}^{gg}$, the cross-correlation $\tilde{C}^{g\kappa}$ and their combination $\tilde{C}^{gg}+\tilde{C}^{g\kappa}$, assuming the redshift distribution \texttt{SKADS}, a linear (denoted with 'lin') and \texttt{HALOFIT} power spectrum (denoted with 'nl'), and fixing the cosmology to the fiducial values. Blue (orange) errorbars correspond to the flux density cut 0.18 (0.4) mJy and solid (dashed) lines to the constant bias model (constant amplitude model). \textit{Right:} Same as in the left panel but now for the $\sigma_8$ constraints on the combined spectra. The bottom lines present the \textit{Planck} \citep{Planck2020}, DES \citep{Abbott21} and KiDS \citep{Heymans2021} measurements with red, magenta and green color, respectively.}
\label{fig:1d_plots}
\end{figure*}

Thus, in the main analysis of \autoref{subsec:Gal_bias_results} and \autoref{subsec:sigma8} 
with the $0.18$ mJy flux density cut, we opt to use the scale range $\ell\in(2, 250)$ for the auto-correlation $gg$\footnote{Alternative ways to deal with the small-scale offset between the \texttt{Selavy} and \texttt{PyBDSF} spectra, are to take the difference of the two and introduce an extra nuisance amplitude parameter added in the data covariance to be marginalised over, or to take the cross spectrum between the two catalogues. Nonetheless, we are being conservative in this work and apply a scale cut, while we leave the other alternatives to be investigated in a future work.} and the full range $\ell\in(2, 500)$ for the cross-correlation $g\kappa$. Also, since \texttt{Selavy} and \texttt{PyBDSF} agree at the scales we mentioned {(as we saw at 1$\sigma$)}, we proceed in the analysis of the main results of \autoref{subsec:Gal_bias_results} and \autoref{subsec:sigma8} using the \texttt{Selavy} catalogue alone. 

We quantify the significance of detection as: $\text{SNR}=\sqrt{\chi^2_\text{null}-\chi^2_{b.f.}}$, in terms of $\sigma$, where $\chi^2_\text{null}$ is the $\chi^2$ of the null hypothesis (zero theory vector) and $\chi^2_\text{b.f.}$ the best-fit model $\chi^2$. Regarding the scale cut at $\ell=250$ for $gg$, most of the signal is at $\ell<250$, since there, we obtain a detection of $11\sigma$, while at the full scale range the detection is $14 \sigma$. The cross-correlation $g\kappa$ detection significance up to $\ell=500$ is $5.5\sigma$.

We repeat the same for the more conservative flux density cut at $0.4$mJy and show the results for the $gg$ in the top right panel of \autoref{fig:scale_cut_flux_0p18}. Now, the catalogues agree with each other at the full scale range up to $\ell=500$ {always within 1$\sigma$}, even though \texttt{PyBDSF} has again slightly {($\sim0.5\sigma$)} more power than \texttt{Selavy} at small scales. This is further confirmed by the theoretical models which yield consistent results. Therefore, we opt to use the full scale range for $gg$ and use the \texttt{Selavy} catalogue alone for the galaxy bias and cosmology analysis of \autoref{sec:results}. At this point, we mention that the agreement we see now at the flux density cut $0.4$ mJy between the catalogues can be attributed to the fact that we consider a more conservative galaxy sample which at the same time contains less galaxies (and in turn, larger uncertainties inflating the errorbars) than the sample with flux density cut at $0.18$ mJy (see \autoref{subsubsec:Source finding algorithms}). Regarding, the cross-correlation spectra $g\kappa$ for the flux density cut at $0.4$ mJy, we find very similar results at the whole scale range with those obtained at $0.18$ mJy and therefore we do not show them in the panel to avoid repetition. The cross-correlation detection significance for $0.4$ mJy is $5.4\sigma$

\begin{figure}[hbt!]
\centering
\includegraphics[width=0.5\textwidth]{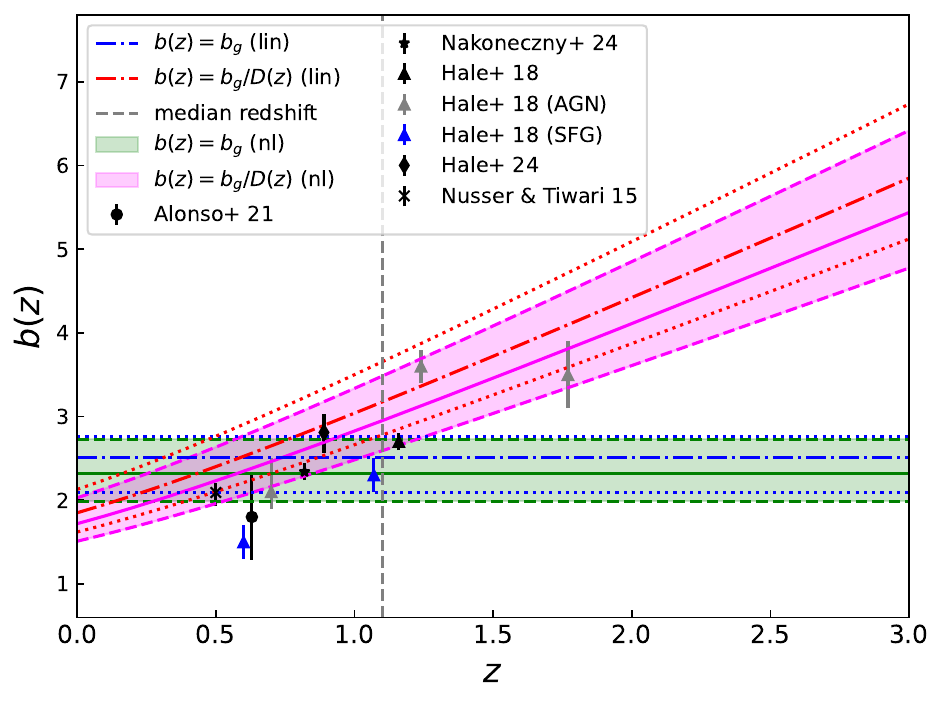}
\caption{Best-fit values along with the 68\% confidence interval constraints on the constant bias (green and blue) and constant amplitude (magenta and red) model for the combined spectra $\tilde{C}^{gg}+\tilde{C}^{g\kappa}$ assuming a \texttt{SKADS} distribution, a \texttt{HALOFIT} (filled intervals) {as well as a linear (empty intervals)} spectrum and a flux density cut at 0.18mJy. The errorbars with the different marker styles represent galaxy bias measurements from different radio galaxy surveys in the literature. Grey and blue triangular markers correspond to AGN and SFG constraints as from H18 \citep{Hale18} while the black triangular marker to the combined sample in the same work. The rest of the different shape black markers show mixed populations from the works \citep{Nusser15, Hale18, Alonso2021, Hale24, Nakoneczny24}.{The vertical dashed line is the median redshift of the sample.}}
\label{fig:biases}
\end{figure}

\subsection{Constraints on galaxy bias}
\label{subsec:Gal_bias_results}

By using the measurements and scale cuts discussed in \autoref{subsec:detectionsig}, first we present the constraints on the galaxy bias $b_g$ while we fix the cosmological parameters, including $\sigma_8$, to the \cite{Planck2020} best-fit values (see again \autoref{subsec:Matter_PS_Galaxy_Bias_Models}). The results are shown in the left panel of \autoref{fig:1d_plots} and the values are in \autoref{tab:flux_0p18_bg_const.} to \autoref{tab:flux_0p4_bg_Dz.}. For our baseline results here we assume the \texttt{SKADS} distribution. We also repeated the analysis with the \texttt{T-RECS} distribution and the results are in agreement with those using \texttt{SKADS} {finding a shift at most of $\sim0.5\sigma$ for the 0.18mJy flux density cut and the constant galaxy bias (lower galaxy bias values and higher $\sigma_8$ values with \texttt{T-RECS}) and even less for the rest of the cases. Therefore, we show the \texttt{T-RECS} results and how they compare to \texttt{SKADS} ones in detail in \autoref{app:trecs_analysis}}.

In the left panel of \autoref{fig:1d_plots}, we report the constant galaxy bias best fit and 68\% confidence interval model constraints with blue solid lines for the flux density cut at $0.18$ mJy. We find that the auto-correlation gives higher bias than the cross-correlation by $\sim 1\sigma$, while the combination of the two gives intermediate estimates. In all these results the linear model yields higher bias values than \texttt{HALOFIT} by $\sim0.5\sigma$ to compensate for the smaller power at the mildly non-linear regime. Also, we do not observe deviations from the shot noise estimates given the reported constraints on the nuisance amplitude parameter $A_\text{sn}$, verifying in this way that there is no {evidence of} multi-component contamination in the \texttt{Selavy} island catalogue we use. Additionally, it could mean that our low density sample is not affected by other contributions like halo exclusion. Regarding the goodness of fit, we report reduced $\chi^2$ ($\chi^2_\nu$), defined as, $\chi^2_\nu=\chi^2_\text{min}/\nu$, with $\chi^2_\text{min}$ the $\chi^2$ of the best-fit value and $\nu$ the degrees of freedom which is the number of our measurements minus the number of fitted parameters. {We also report the 'probability to exceed' which is defined as $\text{PTE}(\chi^2,\nu)=1-\text{CDF}(\chi^2,\nu)$, where CDF is the cumulative distribution of $\chi^2$.} Overall, all the measurements provide good fits to the data at $\chi^2_\nu \sim 1$ {(or equialently a PTE of 10-90\%)} with the auto-correlation results giving worse $\chi^2_\nu$ than the cross-correlation and the combined ones (see \autoref{tab:flux_0p18_bg_const.}). We opt to report in the text for clarity (and do so for the rest of the galaxy models and flux density cuts in the paragraphs below) only the combined measurements $(gg+g\kappa)$ galaxy bias values from \texttt{SKADS} for the \texttt{HALOFIT} model, which is $b_g=2.32^{+0.41}_{-0.33}$. The rest of the results are shown \autoref{tab:flux_0p18_bg_const.}.

Turing our attention to the constant amplitude model results (see values at \autoref{tab:flux_0p18_bg_Dz.}), which are shown with dashed blue lines of the left panel of \autoref{fig:1d_plots}, the aspects we discussed for the constant galaxy bias model, apply similarly here. However, the constraints on the amplitude parameter are lower than the simple constant bias, as expected, since we now take into account the growth evolution with redshift in the bias model (see again \autoref{subsec:Matter_PS_Galaxy_Bias_Models}). We quote here the combined measurement galaxy bias estimate from \texttt{SKADS} for \texttt{HALOFIT} as $b_g=1.72^{+0.31}_{-0.21}$.

The same picture concerning the differences between the linear and non-linear power spectrum recipes also applies for the constant galaxy bias model constraints at the more conservative flux density cut of $0.4$ mJy (see \autoref{tab:flux_0p4_bg_const.}) presented with the solid orange lines in the left panel of \autoref{fig:1d_plots}. Although in this sample we have fewer radio galaxies than for the flux density cut at $0.18$ mJy (see again \autoref{subsubsec:Source finding algorithms} for the reported number of sources), at the same time at the $0.4$ mJy flux density cut we consider scales up to $\ell=500$ and do not cut at $\ell=250$ as we do for the less conservative cut, gaining in  this way more constraining power. Thus, the resulting constraints shrink by up to $\sim$30\% for the auto-correlation. {It is noteworthy that this increase of the constraining power that we see in the galaxy bias using the brighter and less dense sample (0.4mJy) compared to the fainter one (0.18mJy) may be a critical point for future radio continuum data which consider auto-correlations. This can be clearly seen by the fact that the increase of uncertainty of the power spectrum per multipole can be overcompensated by pushing more towards smaller scales which we can still trust given the agreement between the different catalogues (\texttt{Selavy} and \texttt{PyBDSF}, see again \autoref{fig:scale_cut_flux_0p18}), leading this way to tighter parameter constraints. \footnote{Another point we should mention is that by looking at the top right panel of \autoref{fig:scale_cut_flux_0p18} for the auto-correlation of the brighter sample, we can appreciate that although the shot-noise starts to be more important than the signal at $\ell>200$, the SNR keeps increasing up to $\sim3\sigma$ until $\ell=500$ and as a result further improves the constraining power.}. We should also mention that apart from the larger errorbars in the brighter sample, also the measurements themselves are in better agreement (between \texttt{Selavy} and \texttt{PyBDSF} at $\ell>250$), which could be indicative of a possible mitigation of the source finding problem for brighter and denser radio samples.} Going now back to the results, for this flux density cut, the auto-correlation and the combined measurements galaxy bias estimates are now lower, while the cross-correlation alone estimates are larger than the results with the previous less conservative flux density cut. Here, the reported combined measurement estimate of the galaxy bias from \texttt{SKADS} for \texttt{HALOFIT} reads $b_g=2.18^{+0.17}_{-0.25}$.

Finally, we display the constraints for the constant amplitude galaxy bias at $0.4$ mJy (see \autoref{tab:flux_0p4_bg_Dz.}) which correspond to the dashed orange lines of the left panel of \autoref{fig:1d_plots}. The trends in the results here for the various modelling assumptions are again consistent with the picture seen for the same galaxy model at $0.18$ mJy, with two noteworthy differences. The first is that the auto-correlation constraints remain about the same irrespective of the flux density cut and the combined spectra results give slightly higher bias at the 0.4 mJy cut. Now, the combined data assuming \texttt{SKADS} and \texttt{HALOFIT} give $b_g=1.78^{+0.22}_{-0.15}$.

In \autoref{fig:biases} we show the best-fit and the 68\% confidence intervals on the constant galaxy bias and the constant amplitude model as a function of redshift for our combined spectra with {the linear and \texttt{HALOFIT} power spectrum } and assuming a \texttt{SKADS} redshift distribution at the flux density cut of $0.18$ mJy (the results are also very similar with the other flux density cut at $0.4$ mJy). We also illustrate the bias measurements from different and mixed populations of galaxies found in other radio continuum works in the literature \citep[see caption for details]{Nusser15, Hale18, Alonso2021, Hale24, Nakoneczny24}, though a direct comparison with them is impossible due to the different combinations of spectra assumed as well as the different effective flux density limits. However, the results slightly hint that the our constant amplitude model for the galaxy bias is a better description for the deep radio continuum galaxy data compared to the constant redshift-independent model. {In order to compare the galaxy bias values of the constant galaxy bias and the constant amplitude model we estimate the constant amplitude constraints at the effective redshift of the \texttt{SKADS} and \texttt{T-RECS} distributions and report them as an extra column (see second column of \autoref{tab:flux_0p18_bg_Dz.} and \autoref{tab:flux_0p4_bg_Dz.}). By comparing the results of the constant amplitude model at the effective redshift with the constant one, we see that the values are higher. This is not surprising given that the constant amplitude model accounts for the linear growth factor whose inverse is greater than unity at high redshifts.} 

{Additionally, it is important to note that one would expect a higher galaxy bias value for a brighter sample ($0.4$mJy) than a fainter one ($0.18$mJy) given that the brighter more luminous sources reside in more massive halos which have larger bias. However, it could be that the fainter sample contains more high-redshift sources which naturally have larger galaxy bias. The latter, could explain our findings here and they agree with the results of \cite{Nakoneczny24}. In any case, further studies would be needed to investigate this.} 

{A crucial point here is that similarly to the uncertainties in the redshift distribution, there exist uncertainties on the validity of the linear galaxy bias models we employ (constant galaxy bias and constant amplitude). These stem from the fact that the redshift distribution of radio samples, though broad enough and extending to high redshift with long tails where linearity can be assumed, they still start near redshift zero, where non-linearities enter even in the larger scales (since smaller physical scales of nearby structures look larger in the sky). This should be kept in mind given the slight differences we see when we compare results from the different galaxy bias models and the linear and the non-linear matter power spectrum recipes we use. Nevertheless, we can safely assume that the impact of non-linearities in our analysis is small, a fact that can be supported by the agreement of the constraints within $1\sigma$ (see again \autoref{fig:biases}). In subsequent future works, and as the number density of the EMU sample will increase as well as its sky coverage and constraining power, we will investigate these effects in more detail by considering non-linear galaxy bias models and pushing to smaller angular scales ($\ell>500$).}

\subsection{Constraints on $\sigma_8$}
\label{subsec:sigma8}

Now, we place constraints on the $\sigma_8$ parameter by leaving it free during the fitting, and we do this for the combined measurements $gg+g\kappa$ in order to break the degeneracy between the galaxy bias $b_g$ and $\sigma_8$. We assume for our baseline the \texttt{SKADS} distribution. Also, we consider the different cuts and the models we discussed in \autoref{subsec:Gal_bias_results}. The results are shown from \autoref{tab:flux_0p18_bg_const.} to \autoref{tab:flux_0p4_bg_Dz.} in the row denoted with `$gg+g\kappa (\sigma_8 \text{ free})$' and in the right panel of \autoref{fig:1d_plots}. For completeness, in  \autoref{fig:sigma_8_bias_contours} of the \autoref{app:trecs_analysis}, we present the marginalised posterior contours along with their 68\% and 95\% confidence intervals and the one-dimensional posteriors for the $b_g$, $\sigma_8$ and $A_\text{sn}$.

Overall, the constraints are not competitive due to the low density sample we consider here, the small sky coverage and the large uncertainties in the redshift distribution. Nonetheless, this work serves as a sanity check and a complementary cosmological constraint on one hand and on the other hand, it demonstrates the potential of the full EMU survey and also of other radio continuum galaxy surveys for cosmological studies (similar to \citealt{Alonso2021, Nakoneczny24} and \citealt{Piccirilli22}).

Regarding the results themselves and focusing on the flux density cut at $0.18 $mJy and the constant bias model, we show its constraints with solid blue lines in the right panel of \autoref{fig:1d_plots} (see also top left figure of the contour plot in \autoref{fig:sigma_8_bias_contours}). We report here the best-fit value using \texttt{HALOFIT} and \texttt{SKADS}, which gives $\sigma_8=0.68^{+0.16}_{-0.14}$. The measurements obtained with the linear and the non-linear \texttt{HALOFIT} are in agreement with each other. Also, they agree with the \textit{Planck} \citep{Planck2020}, DES \citep{Abbott21} and KiDS \citep{Heymans_2021} measurements. We note that we observe a trend for lower $\sigma_8$ values, although they are consistent with the other surveys' results at $1\sigma$. The linear model estimates can be found in \autoref{tab:flux_0p18_bg_const.}.

Turning now to the dashed blue lines of the right panel of \autoref{fig:1d_plots}, we see the constant amplitude model constraints at the same flux density cut. Here the $\sigma_8$ estimates are sightly lower compared to the constant bias results enhancing marginally the preference for lower $\sigma_8$, but again remain consistent within $1\sigma$ with the estimates from the other surveys. The result for \texttt{HALOFIT} is $\sigma_8=0.61^{+0.18}_{-0.20}$. The exact values for the rest of the models are shown in \autoref{tab:flux_0p18_bg_Dz.} and the marginalised contours are presented in the top right panel of \autoref{fig:sigma_8_bias_contours}.

Concerning the results of the constant galaxy bias model at $0.4$ mJy, these are shown with solid orange lines in the right panel of \autoref{fig:1d_plots} (see \autoref{tab:flux_0p4_bg_const.} and bottom left panel of \autoref{fig:sigma_8_bias_contours} for the results of all the models). Now, the most striking difference compared to the results obtained at $0.18$ mJy, is that the $\sigma_8$ is higher, although still consistent at 1$\sigma$. We report the $\sigma_8$ result with \texttt{HALOFIT} which is $0.82\pm0.10$, centered on the \textit{Planck} result. 

Finally, the picture for the constant amplitude model at the $0.4$ mJy flux density cut is as expected and consistent with what we described for the variety of models before. These constraints are shown with dashed orange lines in the right panel of \autoref{fig:1d_plots}. The \texttt{SKADS} result for \texttt{HALOFIT} is $\sigma_8=0.78^{+0.11}_{-0.09}$ (linear model values in \autoref{tab:flux_0p4_bg_Dz.} and the variety of models again shown in the bottom right panel of \autoref{fig:sigma_8_bias_contours} ). {The higher estimated value and the increased constraining power for $\sigma_8$ when a more conservative flux density cut is applied are related to the opposite behavior of the galaxy bias and the reasons we discussed, respectively in \autoref{subsec:Gal_bias_results}}. The higher galaxy bias values for brighter samples has also been seen by \cite{Nakoneczny24}, however, we cannot make a direct comparison with their work, since it concerned an analysis using a denser radio continuum sample of the LoTSS survey \citep{Shimwell_2019} and also applied different flux density cuts.

{Another important point, is the performance of \texttt{T-RECS} compared to the \texttt{SKADS}. Although we discuss the comparison between them in detail in \autoref{app:trecs_analysis}, it is interesting to mention here, that similarly to what we saw for the galaxy bias also applies on $\sigma_8$ but with an opposite behavior. By looking at the right panel of \autoref{fig:1d_plots_trecs}, it is evident that the different redshift distributions can yield up to $\sim0.5\sigma$ parameter shift, assuming a constant galaxy bias model and the 0.18mJy flux density cut.}

\section{Conclusions}
\label{sec:conclusions}
In this work we measured the galaxy clustering auto-correlation harmonic-space power spectrum of the EMU PS1 data and its cross-correlation with the CMB lensing convergence from \textit{Planck} PR4. Then we used these spectra in order to place constraints on the galaxy bias $b_g$ and the amplitude of matter fluctuations $\sigma_8$. 

We studied this for a variety of models. First, we included in our theoretical modelling a linear and a non-linear matter power spectrum using \texttt{HALOFIT} and linear to mildly non-linear scale range from $\ell=2$ to $500$. We also used the redshift distribution from the \texttt{SKADS} simulation in our baseline analysis since we found that using the \texttt{T-RECS} distribution gives consistent results. Then, we assumed two galaxy bias models: a constant redshift-independent galaxy bias $b(z)=b_g$, and a constant amplitude galaxy bias $b(z)=b_g/D(z)$, with $D(z)$ the linear growth factor, which accounts for the redshift evolution of the clustering of radio galaxies. 

For the data, we considered a flux density cut at $0.18$ mJy and an alternative more strict cut at $0.4$ mJy, while we used the \texttt{Selavy} and \texttt{PyBDSF} as our source detection algorithms. After we constructed the weight maps for the two catalogues and propagated their effect in a pseudo-$C_\ell$ analysis, we fitted our data with our theory models in order to put constraints on $b_g$ and $\sigma_8$. This was achieved with an MCMC analysis and using a Gaussian covariance matrix. Below we summarise the most important results we found:
\begin{itemize}
    \item The auto-correlation spectra for EMU PS1 using the \texttt{Selavy} and \texttt{PyBDSF} detection algorithms start to deviate significantly (more than our measurement errors) for $\ell \gtrsim 250$ at the $0.18$ mJy flux density cut. This is due to the fact that the former algorithm categorises large structures as single objects, while the latter categorises possible sub-structures near a main object as different objects. This way, more detected sources lead to a higher power spectrum as measured by \texttt{PyBDSF}. Nonetheless, since both algorithms could be right or wrong on this aspect (finding false negatives or false positives), we ignore scales above 250. To account for any residual uncertainty remaining on the number of sources, we add an extra shot-noise parameter $A_\text{sn}$ in our modelling. We did not report any difference for the $0.4$ mJy flux density cut where we kept all the scale range {a fact that resulted later in increasing the constraining power by $\sim 30\%$ on galaxy bias and $\sigma_8$}.
    \item We found a $\sim$5.5 $\sigma$ detection between the EMU PS1 and the CMB lensing independent of flux density cut.
    \item At the scale regime where our algorithms agreed, we chose \texttt{Selavy} for our baseline analysis and we placed constraints on the galaxy bias by fixing the cosmological parameters using auto-correlation, cross-correlation spectra and their combination. All the different models and flux density cuts yield consistent results. {We found that there is a shift of $\sim0.5\sigma$ depending on the linear and non-linear \texttt{HALOFIT} matter power spectrum which is a systematic effect, related to the non-linear galaxy bias modelling.} Assuming a \texttt{HALOFIT} model and the $0.18$ mJy ($0.4$ mJy) flux density cut on the combined spectra, we report a constant galaxy bias of $b_g=2.32^{+0.41}_{-0.33}$ ($b_g=2.18^{+0.17}_{-0.25}$) and a constant amplitude galaxy bias of $b_g=1.72^{+0.31}_{-0.21}$ ($b_g=1.78^{+0.22}_{-0.15}$).
    \item After freeing $\sigma_8$ for the same theory model and flux density cut choices we found $\sigma_8=0.68^{+0.16}_{-0.14}$ for the constant bias and $\sigma_8=0.61^{+0.18}_{-0.20}$ for the constant amplitude model. These values increase slightly for the $0.4$ mJy flux density cut at $\sigma_8=0.82\pm0.10$ and $\sigma_8=0.78^{+0.11}_{-0.09}$, respectively. These values are in very good agreement with the \textit{Planck} CMB measurements \citep{Planck2020}, and the weak lensing surveys of Dark Energy Survey (DES; \citealt{Abbott21}) and Kilo Degree Survey (KiDS; \citealt{Heymans_2021}).
\end{itemize}

This paper highlights the possibility to break the degeneracy between the galaxy bias $b_g$ and the amplitude of the matter fluctuations $\sigma_8$ by using auto-correlation and cross-correlation of the radio continuum galaxy sample from the EMU PS1 with the CMB lensing convergence as from \textit{Planck} PR4. The largest bottleneck for the deep radio continuum samples remains the insufficient information on their redshift distribution. There is ongoing work on the cross-correlation of EMU PS1 data with optical galaxies from DES dealing with the modelling of the redshift distribution {\citep{Saraf_2025}}. In coming years there will be more data covering a larger fraction of the sky ($\sim$ 50\%), which will certainly reduce the uncertainties on the galaxy bias and the cosmological parameters. In addition, large optical surveys like Euclid can help in reducing the uncertainties on the redshift estimates of the radio galaxy sample {\citep{Kalus_2025}}. This will be achieved by cross-matching radio galaxies with their optical counterparts at known redshifts. This work is only a first step of what EMU survey can achieve even with the pilot survey data covering a relatively small and contiguous patch of sky ($\sim 270$ deg$^{2}$). Eventually, by combining its deep observations with the large sky area, EMU will manage to probe the matter distribution of the large-scale structure at huge volumes, which will be ideal for studies on extensions to the $\Lambda$CDM model \citep[{e.g.}][]{Alonso_2015, Camera2012, Bernal_2019}.

\section*{Acknowledgements}
The authors thank David Alonso for the enlightening discussions and feedback which were invaluable for the realisation of this project. We also thank the anonymous referee for their constructive comments which improved the presentation of this manuscript. KT is supported by the STFC grant ST/W000903/1 and by
the European Structural and Investment Fund. KT also acknowledges the use of the PHOEBE computing facility in Czech Republic. JA is supported by MICIIN-European Union NextGenerationEU Mar\'ia Zambrano program (UCM CT33/21), the UCM project PR3/23-30808, MICINN (Spain) grant PID2022-138263NB-I0 (AEI/FEDER, UE) and the Diputaci\'on General de Arag\'on-Fondo Social Europeo (DGA-FSE) Grant No. 2020-E21-17R of the Aragon Government. JA acknowledges the use of Aljuarismi UCM computing facility. CLH acknowledges
generous support from the Hintze Family Charitable Foundation
through the Oxford Hintze Centre for Astrophysical Surveys. BB-K acknowledges support from INAF for the project "Paving the way to radio cosmology in the SKA Observatory era: synergies between SKA pathfinders/precursors and the new generation of optical/near-infrared cosmological surveys" (CUP C54I19001050001). SC acknowledges support from the Italian Ministry of University and Research (\textsc{mur}), PRIN 2022 `EXSKALIBUR – Euclid-Cross-SKA: Likelihood Inference Building for Universe's Research', Grant No.\ 20222BBYB9, CUP D53D2300252 0006, from the Italian Ministry of Foreign Affairs and International
Cooperation (\textsc{maeci}), Grant No.\ ZA23GR03, and from the European Union -- Next Generation EU.
MB is supported by the Polish National Science Center through grants no. 2020/38/E/ST9/00395, 2018/30/E/ST9/00698, 2018/31/G/ST9/03388 and 2020/39/B/ST9/03494.

The Australian SKA Pathfinder is part of the Australia Telescope National Facility (\hyperlink{https://ror.org/05qajvd42}{https://ror.org/05qajvd42}) which is managed by CSIRO. Operation of ASKAP is funded by the Australian Government with support from the National Collaborative Research Infrastructure Strategy. ASKAP uses the resources of the Pawsey Supercomputing Centre. Establishment of ASKAP, the Murchison Radio-astronomy Observatory and the Pawsey Supercomputing Centre are initiatives of the Australian Government, with support from the Government of Western Australia and the Science and Industry Endowment Fund. We acknowledge the Wajarri Yamatji people as the traditional owners of the Observatory site.






\bibliographystyle{mnras}
\bibliography{biblio} 




\appendix

\section{Quadratic galaxy bias model}
\label{app:quadratic_model}

\begin{figure}[hbt!]
\centering
\includegraphics[width=0.435\textwidth]{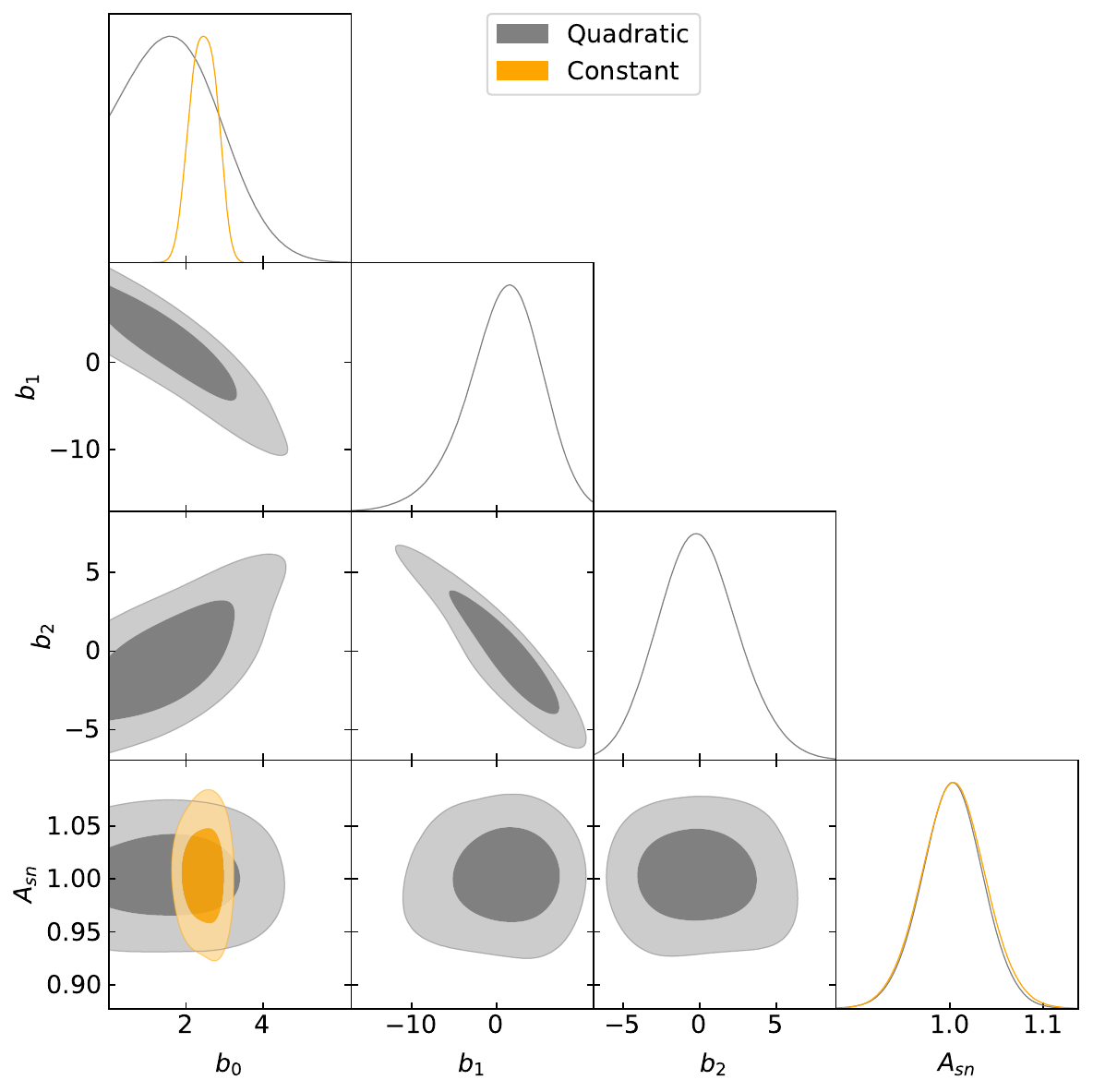}
\caption{The 68\% and 95\% confidence intervals of the marginalised contours and the one-dimensional posteriors for the galaxy bias parameters of the quadratic (grey), the constant model (orange) and $A_\text{sn}$. Also, the \texttt{Selavy} catalogue was used and we considered a flux density cut at 0.18mJy. We assumed a fixed cosmology, an \texttt{SKADS} distribution and a \texttt{HALOFIT} power spectrum.}
\label{fig:quadratic}
\end{figure}

Here, we test the quadratic galaxy bias model, which as we saw in \autoref{subsec:Matter_PS_Galaxy_Bias_Models} is a polynomial function in redshift $b(z)=b_0+b_1z+b_2z^2$ that has three free parameters: $b_0$, $b_1$ and $b_2$. We assume a positive flat prior for $b_0$, while we allow both negative and positive values for the flat priors of $b_1$ and $b_2$, which correspond to the redshift dependence and the evolution of the galaxy bias. In \autoref{fig:quadratic} we show the constraints of the quadratic galaxy bias model parameters and keep $\sigma_8$ to its fiducial value. These are shown with the grey contours for which we assume only the auto-correlation $\tilde{C}^{gg}$, a \texttt{HALOFIT} power spectrum, the redshift distribution of \texttt{SKADS} and the flux density cut of $0.18$ mJy. The results are similar, whether we consider a linear power spectrum, a \texttt{T-RECS} distribution or a flux density cut at $0.4$ mJy. On top of these results, we present the constraints of the same fiducial case for the constant galaxy bias model, which are shown with the orange contour. We can appreciate that the results we obtain with the considerably less constraining quadratic model are consistent with those from the constant galaxy bias. That is, the parameter $b_0$, while poorly constrained, agrees with the $b_g$ estimate of the simple model and the other higher order parameters ($b_1,b_2$) are consistent with the zero value. 

These results mean that the data are not constraining enough to show any other preference for the galaxy bias apart from the single parameter models. This is also verified by the large value in the goodness-of-fit, which is found to be $\chi^2_\nu \sim 3$. Therefore, we opt not to consider the quadratic model for the fiducial analysis in \autoref{sec:results}.

Finally, we note that another possibility could be to use an intermediate linear model $b(z)=b_0+b_1z$ which is empirical, but we do not opt to do so here, since we still have an extra parameter compared to the constant bias and the constant amplitude models we use in our fiducial analysis which have only a single parameter. However, we plan to explore this intermediate model in a future work using more data that will improve our constraining power.

\section{Comparison with sample covariance}
\label{app:mocks_cov}






In order to test the robustness of our fiducial analysis against the analytical covariance of \autoref{eq:covs}, we compare our pipeline with one set of models using the sample covariance. We construct $1000$ mock realisations of correlated \textit{Planck} PR4 CMB convergence and galaxy density field using \texttt{GLASS} \citep{Tessore2023GLASS:}. The galaxy number density and survey area in simulations are consistent with EMU PS1 (see \autoref{subsec:EMUPS}). The redshifts of the simulated set of galaxies follow the \texttt{SKADS} redshift distribution with flux density cut at $0.18$ mJy.\\

\begin{figure}[hbt!]
\centering
\includegraphics[width=0.435\textwidth]{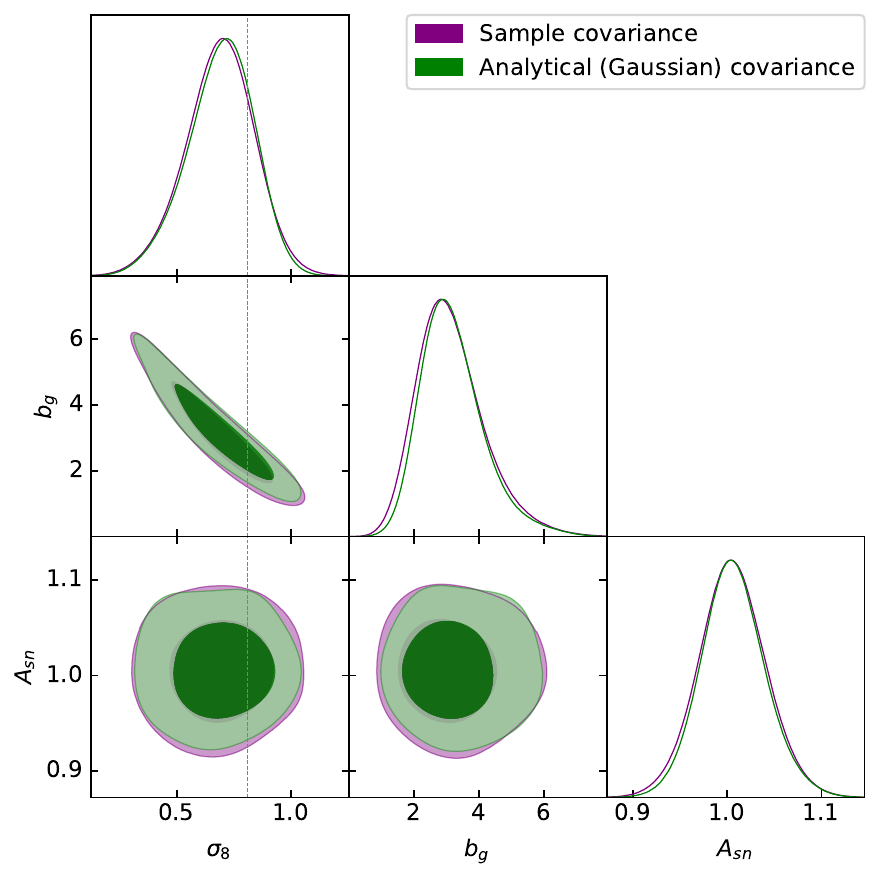}
\caption{Similar to \autoref{fig:quadratic} but only for the constant galaxy bias model between the analytical Gaussian covariance (green) and the sample covariance (purple). The vertical dashed black line marks the best-fit value as from \citealt{Planck2020}.}
\label{fig:covs_comparison}
\end{figure}

\begin{figure*}[tbh!]
\centering
\includegraphics[width=0.55\textwidth]{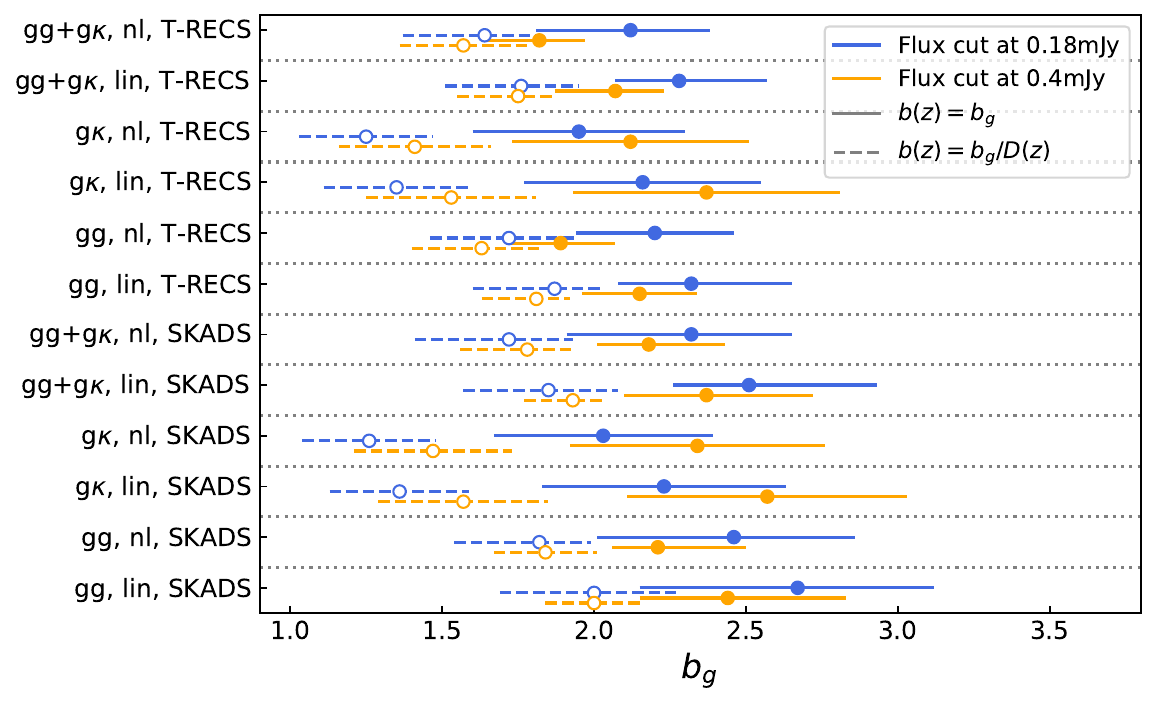}\includegraphics[width=0.5\textwidth]{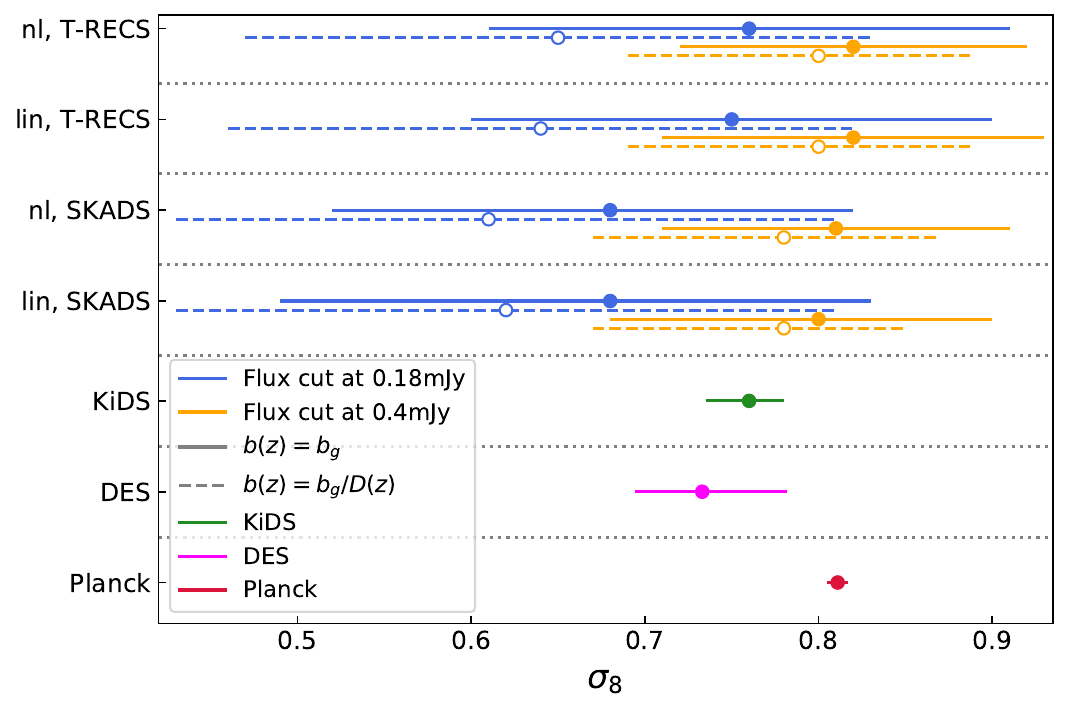}
\caption{Same as \autoref{fig:1d_plots} but now showing on top also the constraints using the \texttt{T-RECS} redshift distribution.}
\label{fig:1d_plots_trecs}
\end{figure*}

We compute the pseudo-$C_\ell$ power spectra from simulations with the same method described in \autoref{subsec:pseudo-cl}. Then, we construct the sample covariance as
\begin{multline}
\mathbf{K}^{WX,YZ}_{\ell\ell'}=\frac{1}{N_{m}-1} \\
\times \sum_{m=1}^{N_m}\left(\tilde{C}^{WX,m}_\ell-\left\langle \tilde{C}^{WX}_{\ell} \right \rangle\right)\left(\tilde{C}^{YZ,m}_\ell-\left\langle  \tilde{C}^{YZ}_{\ell}\right\rangle\right),
\label{eq:mocks_cov}
\end{multline}
where $N_{m}$ is the total number of simulations, $\tilde{C}^{WX,m}_\ell$ is the power spectrum estimated from the $m\text{th}$ simulation and
\begin{equation}
    \left\langle \tilde{C}^{WX}_{\ell} \right\rangle= \frac{1}{N_{m}}\sum_{m=1}^{N_m}\tilde{C}^{WX,m}_\ell.
\label{eq:mock_avg}
\end{equation}

Although the numerical covariance can be an unbiased estimator of the true covariance, its inverse is not and a correction must be applied, known as the \textit{Anderson-Hartlap} factor \citep{Anderson2004, Hartlap2007} which is
\begin{equation}
\mathbf{K}^{-1} \rightarrow{\frac{N_m-N_d-2}{N_m-2}\mathbf{K}^{-1}},
    \label{eq:Hartlap_corr}
\end{equation}

with $N_d$ the size of the data vector. We find excellent agreement between the results using the analytical Gaussian covariance of \autoref{eq:covs} with \texttt{HALOFIT} and the sample covariance of \autoref{eq:mocks_cov} for the same fiducial cosmological model. These are shown in \autoref{fig:covs_comparison}.

\begin{figure*}[tbh!]
\centering
\includegraphics[width=0.5\textwidth]{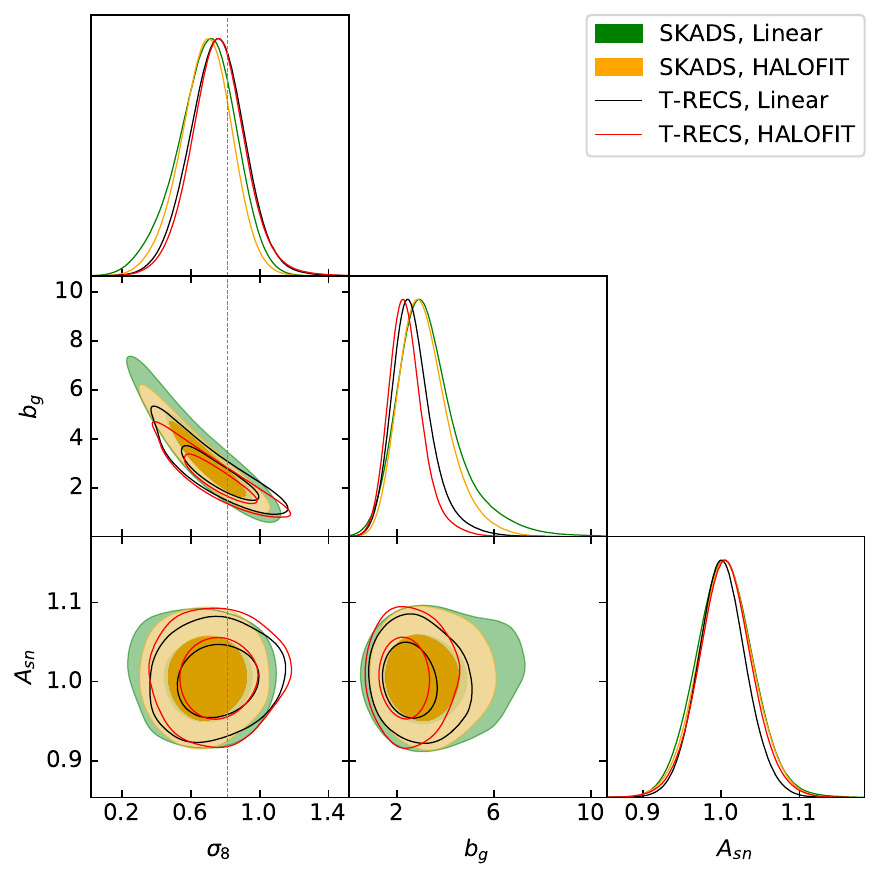}\includegraphics[width=0.5\textwidth]{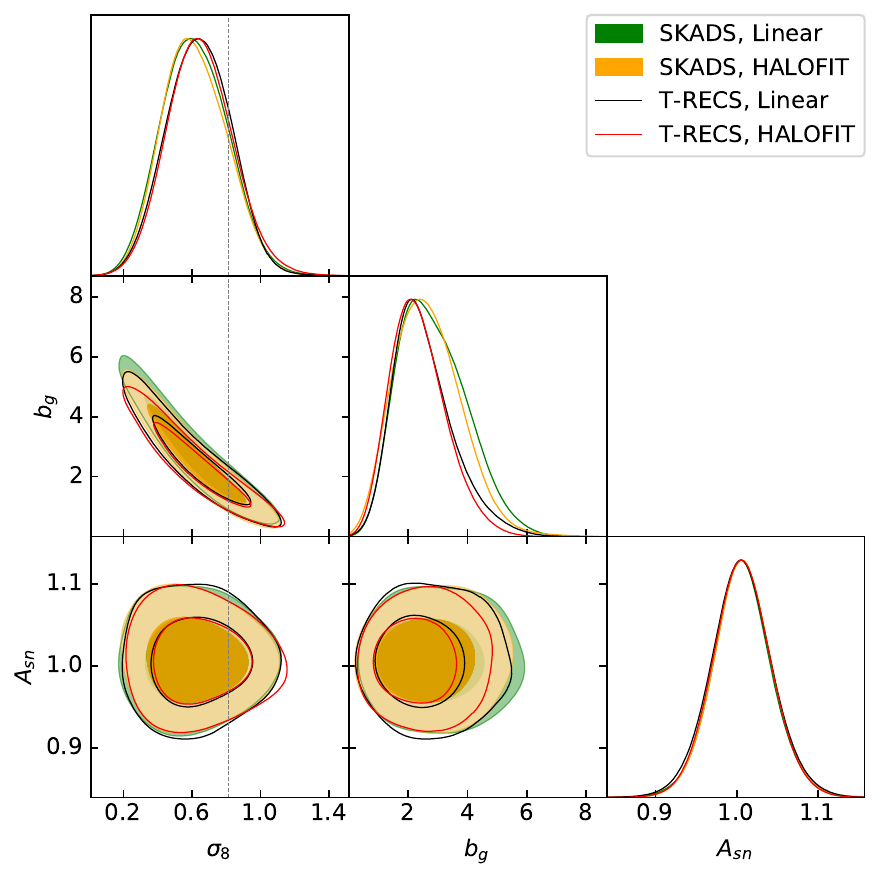}\\
\includegraphics[width=0.5\textwidth]{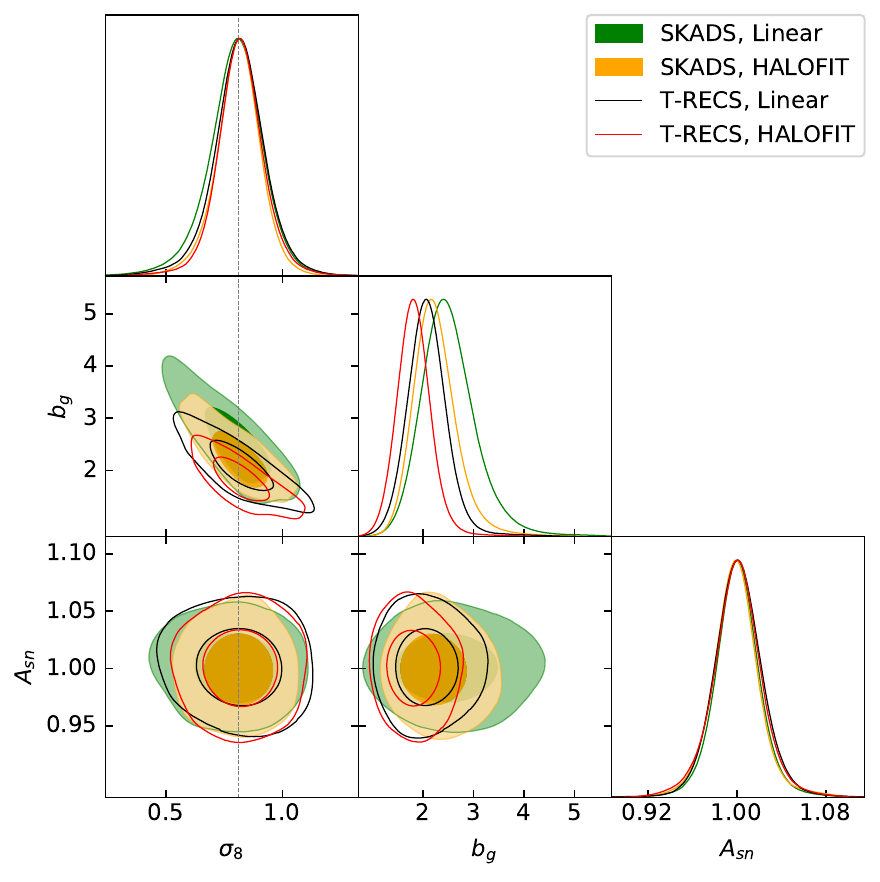}\includegraphics[width=0.5\textwidth]{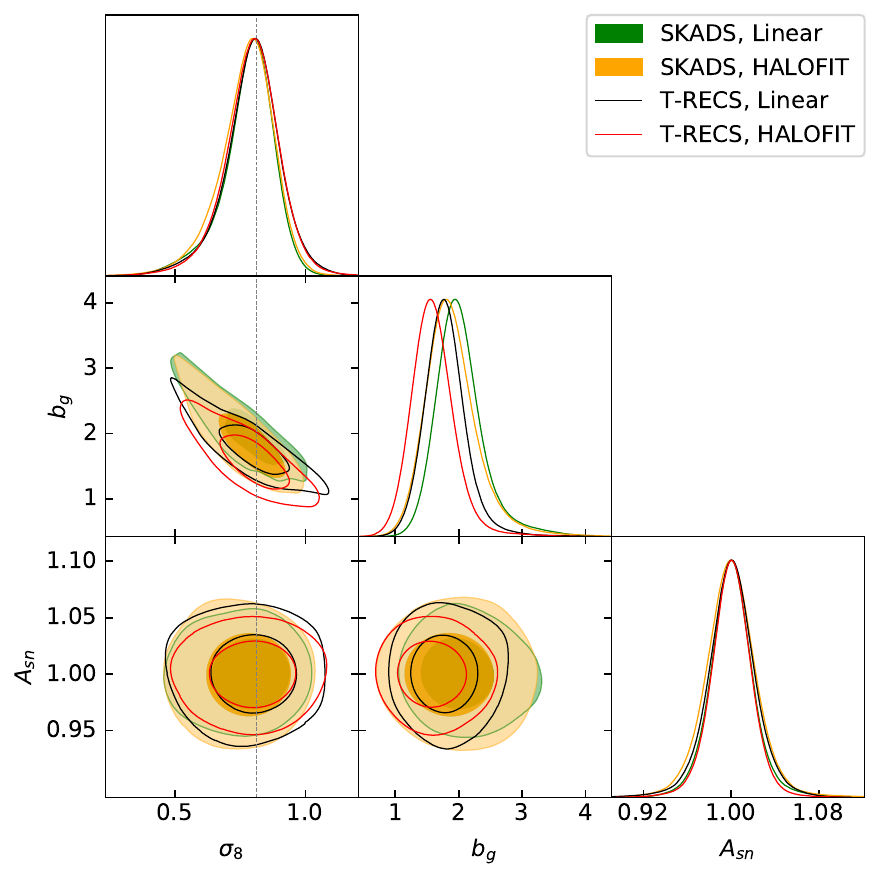}
\caption{The 68\% and 95\% confidence intervals of the marginalised contours and the one-dimensional posteriors for the parameters $b_g$, $\sigma_8$ and $A_\text{sn}$ for the \texttt{Selavy} catalogue. Top and bottom panels show the results for the flux density cut of 0.18mJy and 0.4mJy, while left and right panels correspond to the constant galaxy bias and the constant amplitude model, respectively. Filled contours show constraints from \texttt{SKADS} and empty contours from \texttt{T-RECS}. Cold colours (black and green) denote the linear and warm colours (red and orange) the \texttt{HALOFIT} power spectrum. Again, we remind the reader that the vertical dashed black line marks the best-fit value as from \citealt{Planck2020}. }
\label{fig:sigma_8_bias_contours}
\end{figure*}

\section{Analysis with the T-RECS redshift distribution and comparison with SKADS}
\label{app:trecs_analysis}

As we can see in \autoref{fig:1d_plots_trecs} and \autoref{fig:sigma_8_bias_contours}, the galaxy bias and $\sigma_8$ results obtained with the \texttt{T-RECS} redshift distribution are in total agreement (very well within 1$\sigma$) with those obtained in our baseline analysis with the \texttt{SKADS} which we discussed in \autoref{subsec:Gal_bias_results} and \autoref{subsec:sigma8}. In addition, we can appreciate that the interplay between the different flux density cuts, the galaxy bias models and the linear and non-linear \texttt{HALOFIT} matter power spectra also applies for the \texttt{T-RECS}. However, there is some systematic noticeable difference compared to the results obtained with \texttt{SKADS}.

Regarding the galaxy bias constraints at fixed cosmology, we note that for all the models with the \texttt{T-RECS} redshift distribution give lower galaxy bias results than \texttt{SKADS} to compensate for its higher power. This is explained, as we already mentioned in \autoref{subsubsec:red_dist}, due to the longer high-$z$ tail of the \texttt{SKADS}, and the more localized distribution of the \texttt{T-RECS}. Again, we choose to report here only the values of the constant galaxy bias model for the combined measurements $(gg+g\kappa)$ of \texttt{T-RECS} with the \texttt{HALOFIT} power spectrum. For the rest of the models we refer the reader again to see from \autoref{tab:flux_0p18_bg_const.} to \autoref{tab:flux_0p4_bg_Dz.}. For the flux density cut at $0.18$ mJy, the constant model gives $b_g=2.12^{+0.31}_{-0.26}$, while for the constant amplitude model we find $1.64^{+0.27}_{-0.17}$. As for the flux density cut at $0.4$ mJy, \texttt{T-RECS} gives a constant bias of $b_g=1.82^{+0.18}_{-0.15}$ and a constant amplitude bias of $1.57\pm0.21$.

\renewcommand{\arraystretch}{1.8}
\begin{table*}[tbh!]
\centering
\caption{Summary of the best-fit values and their 68\% confidence intervals for the constant galaxy bias parameter $b_g$, the amplitude shot noise parameter $A_\text{sn}$ and the cosmological parameter $\sigma_8$, at the flux density cut of 0.18mJy. {The last two columns show the $\chi^2_\nu$ and the PTE, respectively}. These results concern $gg$, $g\kappa$ and their combination $gg+g\kappa$ assuming the redshift distributions \texttt{SKADS} and \texttt{T-RECS} and the linear and \texttt{HALOFIT} matter power spectrum. (denoted in the table with `lin' and `nl', respectively.)}
\begin{tabular}{lll|lllll|ll}
\cline{4-8}
\multicolumn{3}{l|}{\multirow{2}{*}{}}                                                                  & \multicolumn{5}{l|}{$\qquad \qquad b(z)=b_g$ at flux density cut 0.18mJy}                                                           &  &  \\ \cline{4-8}
\multicolumn{3}{l|}{}                                                                                   & \multicolumn{1}{l|}{$\qquad b_g$} & \multicolumn{1}{l|}{$\qquad A_\text{sn}$} & \multicolumn{1}{l|}{$\qquad\sigma_8$} & \multicolumn{1}{l|}{$\chi^2_\nu$} &  PTE &  &  \\ \cline{1-8}
\multicolumn{1}{|l|}{\multirow{8}{*}{SKADS}}  & \multicolumn{1}{l|}{\multirow{2}{*}{$\qquad \quad gg$}}         & lin & \multicolumn{1}{l|}{$2.67\substack{+0.52 \\ -0.45}$}   & \multicolumn{1}{l|}{$0.99\pm 0.03$}    & \multicolumn{1}{l|}{}       &  \multicolumn{1}{l|}{1.72} & 14\%   &  &  \\ \cline{3-8}
\multicolumn{1}{|l|}{}                        & \multicolumn{1}{l|}{}                             & nl  & \multicolumn{1}{l|}{$2.46\substack{+0.45 \\ -0.40}$}   & \multicolumn{1}{l|}{$1.00\pm 0.03$}    & \multicolumn{1}{l|}{}       &   \multicolumn{1}{l|}{1.66}&  16\%  &  &  \\ \cline{2-8}
\multicolumn{1}{|l|}{}                        & \multicolumn{1}{l|}{\multirow{2}{*}{$\qquad \quad g\kappa$}}         & lin & \multicolumn{1}{l|}{$2.23\pm{0.40}$}   & \multicolumn{1}{l|}{}    & \multicolumn{1}{l|}{}       &  \multicolumn{1}{l|}{1.11} &  35\%  &  &  \\ \cline{3-8}
\multicolumn{1}{|l|}{}                        & \multicolumn{1}{l|}{}                             & nl  & \multicolumn{1}{l|}{$2.03\pm{0.36}$}   & \multicolumn{1}{l|}{}    & \multicolumn{1}{l|}{}       &  \multicolumn{1}{l|}{1.07} &  38\%  &  &  \\ \cline{2-8}
\multicolumn{1}{|l|}{}                        & \multicolumn{1}{l|}{\multirow{2}{*}{$\qquad gg+g\kappa$}}      & lin & \multicolumn{1}{l|}{$2.51\substack{+0.25 \\ -0.42}$}   & \multicolumn{1}{l|}{$1.00\pm0.03$}    & \multicolumn{1}{l|}{}       &  \multicolumn{1}{l|}{0.92} &  54\%  &  &  \\ \cline{3-8}
\multicolumn{1}{|l|}{}                        & \multicolumn{1}{l|}{}                             & nl  & \multicolumn{1}{l|}{$2.32\substack{+0.41 \\ -0.33}$}   & \multicolumn{1}{l|}{$1.00\pm0.03$}    & \multicolumn{1}{l|}{}       &  \multicolumn{1}{l|}{0.96} &  49\%  &  &  \\ \cline{2-8}
\multicolumn{1}{|l|}{}                        & \multicolumn{1}{l|}{\multirow{2}{*}{$gg+g\kappa(\sigma_8\text{ free})$}}   & lin & \multicolumn{1}{l|}{$3.29\substack{+0.90 \\ -1.40}$}   & \multicolumn{1}{l|}{$1.01\pm0.04$}    & \multicolumn{1}{l|}{$0.68\substack{+0.19 \\ -0.15}$}       &  \multicolumn{1}{l|}{1.58} &  7.6\%  &  &  \\ \cline{3-8}
\multicolumn{1}{|l|}{}                        & \multicolumn{1}{l|}{}                             & nl  & \multicolumn{1}{l|}{$3.10\substack{+0.81 \\ -1.20}$}   & \multicolumn{1}{l|}{$1.00\pm0.04$}    & \multicolumn{1}{l|}{$0.68\substack{+0.16 \\ -0.14}$}       &  \multicolumn{1}{l|}{1.34} &  17\%  &  &  \\ \cline{1-8}
\multicolumn{1}{|l|}{\multirow{8}{*}{T-RECS}} & \multicolumn{1}{l|}{\multirow{2}{*}{$\qquad \quad gg$}}         & lin & \multicolumn{1}{l|}{$2.32\substack{+0.24 \\ -0.33}$}   & \multicolumn{1}{l|}{$1.00\pm0.03$}    & \multicolumn{1}{l|}{}       &  \multicolumn{1}{l|}{1.78} &  13\%  &  &  \\ \cline{3-8}
\multicolumn{1}{|l|}{}                        & \multicolumn{1}{l|}{}                             & nl  & \multicolumn{1}{l|}{$2.20\pm0.26$}   & \multicolumn{1}{l|}{$1.00\pm0.03$}    & \multicolumn{1}{l|}{}       &  \multicolumn{1}{l|}{1.74} &  14\%  &  &  \\ \cline{2-8}
\multicolumn{1}{|l|}{}                        & \multicolumn{1}{l|}{\multirow{2}{*}{$\qquad \quad g\kappa$}}         & lin  & \multicolumn{1}{l|}{$2.16\pm0.39$}   & \multicolumn{1}{l|}{}    & \multicolumn{1}{l|}{}       &  \multicolumn{1}{l|}{1.19} &  29\%  &  &  \\ \cline{3-8}
\multicolumn{1}{|l|}{}                        & \multicolumn{1}{l|}{}                             & nl  & \multicolumn{1}{l|}{$1.95\pm0.35$}   & \multicolumn{1}{l|}{}    & \multicolumn{1}{l|}{}       &  \multicolumn{1}{l|}{1.12} &  34\%  &  &  \\ \cline{2-8}
\multicolumn{1}{|l|}{}                        & \multicolumn{1}{l|}{\multirow{2}{*}{$\qquad gg+g\kappa$}}      & lin & \multicolumn{1}{l|}{$2.28\substack{+0.21 \\ -0.29}$}   & \multicolumn{1}{l|}{$0.99\pm0.03$}    & \multicolumn{1}{l|}{}       &   \multicolumn{1}{l|}{1.20} &  26\%  &  &  \\ \cline{3-8}
\multicolumn{1}{|l|}{}                        & \multicolumn{1}{l|}{}                             & nl  & \multicolumn{1}{l|}{$2.12\substack{+0.31 \\ -0.26}$}   & \multicolumn{1}{l|}{$1.00\pm0.03$}    & \multicolumn{1}{l|}{}       &   \multicolumn{1}{l|}{1.15} &  30\%  &  &  \\ \cline{2-8}
\multicolumn{1}{|l|}{}                        & \multicolumn{1}{l|}{\multirow{2}{*}{$gg+g\kappa(\sigma_8\text{ free})$}} & lin & \multicolumn{1}{l|}{$2.61\substack{+0.65 \\ -0.91}$}   & \multicolumn{1}{l|}{$1.00\pm0.03$}    & \multicolumn{1}{l|}{$0.75\pm0.15$}       &   \multicolumn{1}{l|}{1.43} &  13\%  &  &  \\ \cline{3-8}
\multicolumn{1}{|l|}{}                        & \multicolumn{1}{l|}{}                             & nl  & \multicolumn{1}{l|}{$2.36\substack{+0.58 \\ -0.78}$}   & \multicolumn{1}{l|}{$1.00\pm0.03$}    & \multicolumn{1}{l|}{$0.76\pm0.15$}       &  \multicolumn{1}{l|}{1.37} &  16\%  &  &  \\ \cline{1-8}
\smallskip
\end{tabular}
\label{tab:flux_0p18_bg_const.}
\end{table*}

\begin{table*}[tbh!]
\centering
\caption{Same as \autoref{tab:flux_0p18_bg_const.} but for the constant amplitude galaxy bias model. Note that we also add an extra column that shows the galaxy bias constraints at the effective redshift $z_\text{eff}=\int z n(z) dz / {\int n(z) dz}$ given the \texttt{SKADS} and \texttt{T-RECS} $n(z)$ distributions.}
\begin{tabular}{lll|llllll|ll}
\cline{4-9}
\multicolumn{3}{l|}{\multirow{2}{*}{}}                                                                  & \multicolumn{6}{l|}{$\qquad \qquad b(z)=b_g/{D(z)}$ at flux density cut 0.18mJy}                                                          &  &  \\ \cline{4-9}
\multicolumn{3}{l|}{}                                                                                   & \multicolumn{1}{l|}{$\qquad b_g$} & \multicolumn{1}{l|}{$\quad b(z_\text{eff})$} & \multicolumn{1}{l|}{$\qquad A_\text{sn}$} & \multicolumn{1}{l|}{$\qquad\sigma_8$} & \multicolumn{1}{l|}{$\chi^2_\nu$} & \multicolumn{1}{l|}{PTE} &  &  \\ \cline{1-9}
\multicolumn{1}{|l|}{\multirow{8}{*}{SKADS}}  & \multicolumn{1}{l|}{\multirow{2}{*}{$\qquad \quad gg$}}         & lin & \multicolumn{1}{l|}{$2.00\substack{+0.31 \\ -0.27}$} & \multicolumn{1}{l|}{$\quad 3.75$}   & \multicolumn{1}{l|}{$1.00\pm0.03$}    & \multicolumn{1}{l|}{}       &   \multicolumn{1}{l|}{1.57} &  17\% &  &  \\ \cline{3-9}
\multicolumn{1}{|l|}{}                        & \multicolumn{1}{l|}{}                             & nl  & \multicolumn{1}{l|}{$1.82\substack{+0.28 \\ -0.17}$} & \multicolumn{1}{l|}{$\quad 3.41$}   & \multicolumn{1}{l|}{$1.00\pm0.03$}    & \multicolumn{1}{l|}{}       &  \multicolumn{1}{l|}{1.52} & 19\%  &  &  \\ \cline{2-9}
\multicolumn{1}{|l|}{}                        & \multicolumn{1}{l|}{\multirow{2}{*}{$\qquad \quad g\kappa$}}         & lin & \multicolumn{1}{l|}{$1.36\pm0.23$} & \multicolumn{1}{l|}{$\quad 2.55$}   & \multicolumn{1}{l|}{}    & \multicolumn{1}{l|}{}       &  \multicolumn{1}{l|}{0.96} & 47\%  &  &  \\ \cline{3-9}
\multicolumn{1}{|l|}{}                        & \multicolumn{1}{l|}{}                             & nl  & \multicolumn{1}{l|}{$1.26\pm0.22$} & \multicolumn{1}{l|}{$\quad 2.36$}   & \multicolumn{1}{l|}{}    & \multicolumn{1}{l|}{}       &   \multicolumn{1}{l|}{0.97} & 46\%  &  &  \\ \cline{2-9}
\multicolumn{1}{|l|}{}                        & \multicolumn{1}{l|}{\multirow{2}{*}{$\qquad gg+g\kappa$}}      & lin & \multicolumn{1}{l|}{$1.85\substack{+0.28 \\ -0.23}$}  & \multicolumn{1}{l|}{$\quad 3.47$}  & \multicolumn{1}{l|}{$1.00\pm0.03$}    & \multicolumn{1}{l|}{}       &  \multicolumn{1}{l|}{1.22} &  24\%  &  &  \\ \cline{3-9}
\multicolumn{1}{|l|}{}                        & \multicolumn{1}{l|}{}                             & nl  & \multicolumn{1}{l|}{$1.72\substack{+0.31 \\ -0.21}$} & \multicolumn{1}{l|}{$\quad 3.22$}   & \multicolumn{1}{l|}{$1.00\pm0.03$}    & \multicolumn{1}{l|}{}       &  \multicolumn{1}{l|}{1.36} &  15\%  &  &  \\ \cline{2-9}
\multicolumn{1}{|l|}{}                        & \multicolumn{1}{l|}{\multirow{2}{*}{$gg+g\kappa(\sigma_8\text{ free})$}}   & lin & \multicolumn{1}{l|}{$2.83\substack{+0.94 \\ -1.30}$} & \multicolumn{1}{l|}{$\quad 5.30$}   & \multicolumn{1}{l|}{$1.00\pm0.04$}    & \multicolumn{1}{l|}{$0.62\pm0.19$}       &   \multicolumn{1}{l|}{1.28} &  21\%  &  &  \\ \cline{3-9}
\multicolumn{1}{|l|}{}                        & \multicolumn{1}{l|}{}                             & nl  & \multicolumn{1}{l|}{$2.66\substack{+0.96 \\ -1.30}$} & \multicolumn{1}{l|}{$\quad 4.99$}   & \multicolumn{1}{l|}{$1.00\pm0.03$}    & \multicolumn{1}{l|}{$0.61\substack{+0.18 \\ -0.20}$}       &  \multicolumn{1}{l|}{1.35} &  16\%  &  &  \\ \cline{1-9}
\multicolumn{1}{|l|}{\multirow{8}{*}{T-RECS}} & \multicolumn{1}{l|}{\multirow{2}{*}{$\qquad \quad gg$}}         & lin & \multicolumn{1}{l|}{$1.87\substack{+0.27 \\ -0.16}$}  & \multicolumn{1}{l|}{$\quad 3.31$}  & \multicolumn{1}{l|}{$1.00\pm0.03$}    & \multicolumn{1}{l|}{}       &  \multicolumn{1}{l|}{1.62} &  16\%  &  &  \\ \cline{3-9}
\multicolumn{1}{|l|}{}                        & \multicolumn{1}{l|}{}                             & nl  & \multicolumn{1}{l|}{$1.72\substack{+0.26 \\ -0.23}$} & \multicolumn{1}{l|}{$\quad 3.04$}   & \multicolumn{1}{l|}{$1.00\pm0.03$}    & \multicolumn{1}{l|}{}       &  \multicolumn{1}{l|}{1.49} &  20\%  &  &  \\ \cline{2-9}
\multicolumn{1}{|l|}{}                        & \multicolumn{1}{l|}{\multirow{2}{*}{$\qquad \quad g\kappa$}}         & lin  & \multicolumn{1}{l|}{$1.35\pm0.24$} & \multicolumn{1}{l|}{$\quad 2.39$}   & \multicolumn{1}{l|}{}    & \multicolumn{1}{l|}{}       &  \multicolumn{1}{l|}{1.02} &  42\%  &  &  \\ \cline{3-9}
\multicolumn{1}{|l|}{}                        & \multicolumn{1}{l|}{}                             & nl  & \multicolumn{1}{l|}{$1.25\pm0.22$} & \multicolumn{1}{l|}{$\quad 2.21$}  & \multicolumn{1}{l|}{}    & \multicolumn{1}{l|}{}       &   \multicolumn{1}{l|}{1.01} &  43\%  &  &  \\ \cline{2-9}
\multicolumn{1}{|l|}{}                        & \multicolumn{1}{l|}{\multirow{2}{*}{$\qquad gg+g\kappa$}}      & lin & \multicolumn{1}{l|}{$1.76\substack{+0.25 \\ -0.19}$} & \multicolumn{1}{l|}{$\quad 3.12$}   & \multicolumn{1}{l|}{$1.00\pm0.03$}    & \multicolumn{1}{l|}{}       &   \multicolumn{1}{l|}{1.14} &  31\%  &  &  \\ \cline{3-9}
\multicolumn{1}{|l|}{}                        & \multicolumn{1}{l|}{}                             & nl  & \multicolumn{1}{l|}{$1.64\substack{+0.27 \\ -0.17}$} & \multicolumn{1}{l|}{$\quad 2.90$}   & \multicolumn{1}{l|}{$1.00\pm0.03$}    & \multicolumn{1}{l|}{}       &  \multicolumn{1}{l|}{1.30} &  19\%  &  &  \\ \cline{2-9}
\multicolumn{1}{|l|}{}                        & \multicolumn{1}{l|}{\multirow{2}{*}{$gg+g\kappa(\sigma_8\text{ free})$}} & lin & \multicolumn{1}{l|}{$2.50\substack{+0.74 \\ -1.20}$} & \multicolumn{1}{l|}{$\quad 4.43$}   & \multicolumn{1}{l|}{$1.00\pm0.04$}    & \multicolumn{1}{l|}{$0,64\pm0.18$}       &  \multicolumn{1}{l|}{1.30} &  19\%  &  &  \\ \cline{3-9}
\multicolumn{1}{|l|}{}                        & \multicolumn{1}{l|}{}                             & nl  & \multicolumn{1}{l|}{$2.34\substack{+0.74 \\ -1.10}$} & \multicolumn{1}{l|}{$\quad 4.14$}   & \multicolumn{1}{l|}{$1.00\pm0.03$}    & \multicolumn{1}{l|}{$0.65\pm0.18$}       &   \multicolumn{1}{l|}{1.30} &  19\%  &  &  \\ \cline{1-9}
\smallskip
\end{tabular}
\label{tab:flux_0p18_bg_Dz.}
\end{table*}

Focusing on the $b_g-\sigma_8$ results we shall discuss them in slightly more detail here. Starting for the flux density cut at $0.18$ mJy and the constant bias model in the top left panel of \autoref{fig:sigma_8_bias_contours} (and with solid blue lines in the right panel of \autoref{fig:1d_plots_trecs}), we see that the \texttt{SKADS} model gives larger bias estimates compared to the \texttt{T-RECS} model, which was also found to be the case in the fixed cosmology case we discussed in the previous paragraph, Now, in turn, this affects the $\sigma_8$ constraint which has a slightly opposite behavior (being larger) to balance the effect. We can also notice here that the linear model has a similar effect when it is compared to \texttt{HALOFIT} but to a lesser extent. We report again here the best-fit values using \texttt{HALOFIT} and \texttt{SKADS} for comparison, which gave $\sigma_8=0.68^{+0.16}_{-0.14}$, while now \texttt{T-RECS} gives $\sigma_8=0.76\pm0.15$. It is evident that these measurements are in great agreement with each other as well as with the \textit{Planck} best-fit value (vertical dashed black line in the panel), while we observe a trend for lower $\sigma_8$ values in the degeneracy direction of the $b_g-\sigma_8$ space. The linear model estimates can be found in \autoref{tab:flux_0p18_bg_const.}.  

As for the constraints of the constant amplitude model for the same flux density cut (shown with dashed blue lines in the right panel of \autoref{fig:1d_plots_trecs}), we can appreciate a similar behavior between the different models but now at a smaller extent, while the whole contour set (top right panel of \autoref{fig:sigma_8_bias_contours}) moves below and slightly to the left from the contours we found for the constant bias model (top left panel of \autoref{fig:sigma_8_bias_contours}) in the $b_g-\sigma_8$ plane, as expected due to the lower values found for the amplitude bias. Similarly to what we saw with the \texttt{SKADS}, there is a preference for lower $\sigma_8$ values in the degenerate direction and the best-fit value for \texttt{T-RECS} and \texttt{HALOFIT} is $\sigma_8=0.65\pm0.18$, in agreement at $1\sigma$ with \textit{Planck} (also see linear model results are in \autoref{tab:flux_0p18_bg_Dz.}).

Then, the results for the $0.4$ mJy flux density cut and the constant galaxy bias are along the same line concerning the interplay between the linear and non-linear models, similarly to what we showed for \texttt{SKADS}. As we discussed in \autoref{subsec:sigma8}, the most important difference is the fact that in all the cases the galaxy bias values are lower than those at the $0.18$ mJy flux density cut and the $\sigma_8$ constraints are higher. The \texttt{HALOFIT} model estimate for \texttt{T-RECS} is $\sigma_8=0.82\pm0.10$ (again for the results of the linear model see \autoref{tab:flux_0p4_bg_const.} and all the relevant models with the solid orange lines in the right panel of \autoref{fig:1d_plots_trecs}). It is clear that now there is no preference for lower values for $\sigma_8$ and the \textit{Planck} best-fit value is at the centre of our contours.

Finally, the constant amplitude bias values for \texttt{T-RECS} follow the rational of what we have found for \texttt{SKADS} in \autoref{subsec:sigma8}. For \texttt{HALOFIT} we get $\sigma_8=0.80^{+0.11}_{-0.09}$ (see \autoref{tab:flux_0p4_bg_Dz.} and dashed orange lines in the right panel of \autoref{fig:sigma_8_bias_contours}).

\begin{table*}[tbh!]
\centering
\caption{Same as \autoref{tab:flux_0p18_bg_const.} but for the flux density cut of 0.4mJy.}
\begin{tabular}{lll|lllll|ll}
\cline{4-8}
\multicolumn{3}{l|}{\multirow{2}{*}{}}                                                                  & \multicolumn{4}{l|}{$\qquad \qquad b(z)=b_g$ at flux density cut 0.4mJy}                                                          &  &  \\ \cline{4-8}
\multicolumn{3}{l|}{}                                                                                   & \multicolumn{1}{l|}{$\qquad b_g$} & \multicolumn{1}{l|}{$\qquad A_\text{sn}$} & \multicolumn{1}{l|}{$\qquad \sigma_8$} &\multicolumn{1}{l|}{$\chi^2_\nu$} &  PTE &  &  \\ \cline{1-8}
\multicolumn{1}{|l|}{\multirow{8}{*}{SKADS}}  & \multicolumn{1}{l|}{\multirow{2}{*}{$\qquad \quad gg$}}         & lin & \multicolumn{1}{l|}{$2.44\substack{+0.29 \\ -0.39}$}   & \multicolumn{1}{l|}{$1.00\pm0.01$}    & \multicolumn{1}{l|}{}       &   \multicolumn{1}{l|}{1.43} &  16\%  &  &  \\ \cline{3-8}
\multicolumn{1}{|l|}{}                        & \multicolumn{1}{l|}{}                             & nl  & \multicolumn{1}{l|}{$2.21\substack{+0.15 \\ -0.29}$}   & \multicolumn{1}{l|}{$1.00\pm0.01$}    & \multicolumn{1}{l|}{}       &  \multicolumn{1}{l|}{1.24} &  26\%  &  &  \\ \cline{2-8}
\multicolumn{1}{|l|}{}                        & \multicolumn{1}{l|}{\multirow{2}{*}{$\qquad \quad g\kappa$}}         & lin & \multicolumn{1}{l|}{$2.57\pm0.46$}   & \multicolumn{1}{l|}{}    & \multicolumn{1}{l|}{}       &   \multicolumn{1}{l|}{1.25} &  25\%  &  &  \\ \cline{3-8}
\multicolumn{1}{|l|}{}                        & \multicolumn{1}{l|}{}                             & nl  & \multicolumn{1}{l|}{$2.34\pm0.42$}   & \multicolumn{1}{l|}{}    & \multicolumn{1}{l|}{}       &  \multicolumn{1}{l|}{1.23} &  26\%  &  &  \\ \cline{2-8}
\multicolumn{1}{|l|}{}                        & \multicolumn{1}{l|}{\multirow{2}{*}{$\qquad gg+g\kappa$}}      & lin & \multicolumn{1}{l|}{$2.37\substack{+0.27 \\ -0.35}$}   & \multicolumn{1}{l|}{$1.00\pm0.01$}    & \multicolumn{1}{l|}{}       &  \multicolumn{1}{l|}{1.51} &  6.6\%  &  &  \\ \cline{3-8}
\multicolumn{1}{|l|}{}                        & \multicolumn{1}{l|}{}                             & nl  & \multicolumn{1}{l|}{$2.18\substack{+0.17 \\ -0.25}$}   & \multicolumn{1}{l|}{$1.00\pm0.01$}    & \multicolumn{1}{l|}{}       &   \multicolumn{1}{l|}{0.96} &  50\%  &  &  \\ \cline{2-8}
\multicolumn{1}{|l|}{}                        & \multicolumn{1}{l|}{\multirow{2}{*}{$gg+g\kappa(\sigma_8\text{ free})$}}   & lin & \multicolumn{1}{l|}{$2.51\substack{+0.43 \\ -0.58}$}   & \multicolumn{1}{l|}{$1.00\pm0.02$}    & \multicolumn{1}{l|}{$0.80\substack{+0.12 \\ -0.10}$}       &   \multicolumn{1}{l|}{1.28} &  18\%  &  &  \\ \cline{3-8}
\multicolumn{1}{|l|}{}                        & \multicolumn{1}{l|}{}                             & nl  & \multicolumn{1}{l|}{$2.24\substack{+0.35 \\ -0.45}$}   & \multicolumn{1}{l|}{$1.00\pm0.02$}    & \multicolumn{1}{l|}{$0.81\pm0.10$}       &  \multicolumn{1}{l|}{1.13} &   31\%   &  &  \\ \cline{1-8}
\multicolumn{1}{|l|}{\multirow{8}{*}{T-RECS}} & \multicolumn{1}{l|}{\multirow{2}{*}{$\qquad \quad gg$}}         & lin & \multicolumn{1}{l|}{$2.15\pm0.19$}   & \multicolumn{1}{l|}{$1.00\pm0.01$}    & \multicolumn{1}{l|}{}       &  \multicolumn{1}{l|}{1.72} &  7.8\%  &  &  \\ \cline{3-8}
\multicolumn{1}{|l|}{}                        & \multicolumn{1}{l|}{}                             & nl  & \multicolumn{1}{l|}{$1.89\pm0.18$}   & \multicolumn{1}{l|}{$1.00\pm0.01$}    & \multicolumn{1}{l|}{}       &  \multicolumn{1}{l|}{1.27} &  24\%  &  &  \\ \cline{2-8}
\multicolumn{1}{|l|}{}                        & \multicolumn{1}{l|}{\multirow{2}{*}{$\qquad \quad g\kappa$}}         & lin  & \multicolumn{1}{l|}{$2.37\pm0.44$}   & \multicolumn{1}{l|}{}    & \multicolumn{1}{l|}{}       &   \multicolumn{1}{l|}{1.37} &  18\%  &  &  \\ \cline{3-8}
\multicolumn{1}{|l|}{}                        & \multicolumn{1}{l|}{}                             & nl  & \multicolumn{1}{l|}{$2.12\pm0.39$}   & \multicolumn{1}{l|}{}    & \multicolumn{1}{l|}{}       &   \multicolumn{1}{l|}{1.32} &  21\%  &  &  \\ \cline{2-8}
\multicolumn{1}{|l|}{}                        & \multicolumn{1}{l|}{\multirow{2}{*}{$\qquad gg+g\kappa$}}      & lin & \multicolumn{1}{l|}{$2.07\substack{+0.20 \\ -0.16}$}   & \multicolumn{1}{l|}{$1.00\pm0.01$}    & \multicolumn{1}{l|}{}       &   \multicolumn{1}{l|}{1.73} &  2.2\%  &  &  \\ \cline{3-8}
\multicolumn{1}{|l|}{}                        & \multicolumn{1}{l|}{}                             & nl  & \multicolumn{1}{l|}{$1.82\substack{+0.18 \\ -0.15}$}   & \multicolumn{1}{l|}{$1.00\pm0.01$}    & \multicolumn{1}{l|}{}       &   \multicolumn{1}{l|}{1.09} &  35\%  &  &  \\ \cline{2-8}
\multicolumn{1}{|l|}{}                        & \multicolumn{1}{l|}{\multirow{2}{*}{$gg+g\kappa(\sigma_8\text{ free})$}} & lin & \multicolumn{1}{l|}{$2.09\substack{+0.26 \\ -0.31}$}   & \multicolumn{1}{l|}{$1.00\pm0.02$}    & \multicolumn{1}{l|}{$0.82\pm0.11$}       &   \multicolumn{1}{l|}{1.52} &  6.7\%  &  &  \\ \cline{3-8}
\multicolumn{1}{|l|}{}                        & \multicolumn{1}{l|}{}                             & nl  & \multicolumn{1}{l|}{$1.82\substack{+0.23 \\ -0.27}$}   & \multicolumn{1}{l|}{$1.00\pm0.02$}    & \multicolumn{1}{l|}{$0.82\pm0.10$}       &   \multicolumn{1}{l|}{1.22} &  23\%  &  &  \\ \cline{1-8}
\smallskip
\end{tabular}
\label{tab:flux_0p4_bg_const.}
\end{table*}

\begin{table*}[tbh!]
\centering
\caption{Same as \autoref{tab:flux_0p18_bg_Dz.} but for the flux density cut of 0.4mJy.}
\begin{tabular}{lll|llllll|ll}
\cline{4-9}
\multicolumn{3}{l|}{\multirow{2}{*}{}}                                                                  & \multicolumn{6}{l|}{$\qquad \quad b(z)=b_g/{D(z)}$ at flux density cut 0.4mJy}                                                          &  &  \\ \cline{4-9}
\multicolumn{3}{l|}{}                                                                                   & \multicolumn{1}{l|}{$\qquad b_g$} & \multicolumn{1}{l|}{$\quad b(z_\text{eff})$} & \multicolumn{1}{l|}{$\qquad A_\text{sn}$} & \multicolumn{1}{l|}{$\quad\sigma_8$} & \multicolumn{1}{l|}{$\chi^2_\nu$} &  PTE &  &  \\ \cline{1-9}
\multicolumn{1}{|l|}{\multirow{8}{*}{SKADS}}  & \multicolumn{1}{l|}{\multirow{2}{*}{$\qquad \quad gg$}}         & lin & \multicolumn{1}{l|}{$2.00\pm0.16$} & \multicolumn{1}{l|}{$\quad 3.88$}  & \multicolumn{1}{l|}{$1.00\pm0.01$}    & \multicolumn{1}{l|}{}       &   \multicolumn{1}{l|}{0.99} &  44\%  &  &  \\ \cline{3-9}
\multicolumn{1}{|l|}{}                        & \multicolumn{1}{l|}{}                             & nl  & \multicolumn{1}{l|}{$1.84\pm0.17$} & \multicolumn{1}{l|}{$\quad 3.57$}   & \multicolumn{1}{l|}{$1.00\pm0.01$}    & \multicolumn{1}{l|}{}       &   \multicolumn{1}{l|}{0.99}&  44\%  &  &  \\ \cline{2-9}
\multicolumn{1}{|l|}{}                        & \multicolumn{1}{l|}{\multirow{2}{*}{$\qquad \quad g\kappa$}}         & lin & \multicolumn{1}{l|}{$1.57\pm0.28$} & \multicolumn{1}{l|}{$\quad 3.04$}   & \multicolumn{1}{l|}{}    & \multicolumn{1}{l|}{}       &  \multicolumn{1}{l|}{1.08}&   37\%   &  &  \\ \cline{3-9}
\multicolumn{1}{|l|}{}                        & \multicolumn{1}{l|}{}                             & nl  & \multicolumn{1}{l|}{$1.47\pm0.26$} & \multicolumn{1}{l|}{$\quad 2.85$}   & \multicolumn{1}{l|}{}    & \multicolumn{1}{l|}{}       &   \multicolumn{1}{l|}{1.10} &  35\%  &  &  \\ \cline{2-9}
\multicolumn{1}{|l|}{}                        & \multicolumn{1}{l|}{\multirow{2}{*}{$\qquad gg+g\kappa$}}      & lin & \multicolumn{1}{l|}{$1.93\substack{+0.16 \\ -0.11}$} & \multicolumn{1}{l|}{$\quad 3.75$}   & \multicolumn{1}{l|}{$1.00\pm0.01$}    & \multicolumn{1}{l|}{}       &   \multicolumn{1}{l|}{0.93} &  55\%  &  &  \\ \cline{3-9}
\multicolumn{1}{|l|}{}                        & \multicolumn{1}{l|}{}                             & nl  & \multicolumn{1}{l|}{$1.78\substack{+0.22 \\ -0.15}$} & \multicolumn{1}{l|}{$\quad 3.45$}  & \multicolumn{1}{l|}{$1.00\pm0.01$}    & \multicolumn{1}{l|}{}       &  \multicolumn{1}{l|}{1.04} &  41\%  &  &  \\ \cline{2-9}
\multicolumn{1}{|l|}{}                        & \multicolumn{1}{l|}{\multirow{2}{*}{$gg+g\kappa(\sigma_8\text{ free})$}}   & lin & \multicolumn{1}{l|}{$2.01\substack{+0.28 \\ -0.39}$} & \multicolumn{1}{l|}{$\quad 3.90$}   & \multicolumn{1}{l|}{$1.00\pm0.02$}    & \multicolumn{1}{l|}{$0.78\substack{+0.11 \\ -0.07}$}       &   \multicolumn{1}{l|}{0.93} &  54\%  &  &  \\ \cline{3-9}
\multicolumn{1}{|l|}{}                        & \multicolumn{1}{l|}{}                             & nl  & \multicolumn{1}{l|}{$1.89\substack{+0.32 \\ -0.44}$} & \multicolumn{1}{l|}{$\quad 3.67$}  & \multicolumn{1}{l|}{$1.00\pm0.02$}    & \multicolumn{1}{l|}{$0.78\substack{+0.11 \\ -0.09}$}       &  \multicolumn{1}{l|}{1.05} &  40\%  &  &  \\ \cline{1-9}
\multicolumn{1}{|l|}{\multirow{8}{*}{T-RECS}} & \multicolumn{1}{l|}{\multirow{2}{*}{$\qquad \quad gg$}}         & lin & \multicolumn{1}{l|}{$1.81\substack{+0.18 \\ -0.11}$} & \multicolumn{1}{l|}{$\quad 3.14$}  & \multicolumn{1}{l|}{$1.00\pm0.01$}    & \multicolumn{1}{l|}{}       &   \multicolumn{1}{l|}{1.23} &  27\%  &  &  \\ \cline{3-9}
\multicolumn{1}{|l|}{}                        & \multicolumn{1}{l|}{}                             & nl  & \multicolumn{1}{l|}{$1.63\substack{+0.23 \\ -0.19}$} & \multicolumn{1}{l|}{$\quad 2.83$}  & \multicolumn{1}{l|}{$1.00\pm0.01$}    & \multicolumn{1}{l|}{}       &  \multicolumn{1}{l|}{1.83}  &  5.7\%  &  &  \\ \cline{2-9}
\multicolumn{1}{|l|}{}                        & \multicolumn{1}{l|}{\multirow{2}{*}{$\qquad \quad g\kappa$}}         & lin  & \multicolumn{1}{l|}{$1.53\pm0.28$} & \multicolumn{1}{l|}{$\quad 2.66$}  & \multicolumn{1}{l|}{}    & \multicolumn{1}{l|}{}       &  \multicolumn{1}{l|}{1.17} &  30\%  &  &  \\ \cline{3-9}
\multicolumn{1}{|l|}{}                        & \multicolumn{1}{l|}{}                             & nl  & \multicolumn{1}{l|}{$1.41\pm0.25$}  & \multicolumn{1}{l|}{$\quad 2.45$} & \multicolumn{1}{l|}{}    & \multicolumn{1}{l|}{}       &  \multicolumn{1}{l|}{1.17} &  30\%  &  &  \\ \cline{2-9}
\multicolumn{1}{|l|}{}                        & \multicolumn{1}{l|}{\multirow{2}{*}{$\qquad gg+g\kappa$}}      & lin & \multicolumn{1}{l|}{$1.75\substack{+0.20 \\ -0.11}$} & \multicolumn{1}{l|}{$\quad 3.04$}  & \multicolumn{1}{l|}{$1.00\pm0.01$}    & \multicolumn{1}{l|}{}       &  \multicolumn{1}{l|}{1.06} &  38\%  &  &  \\ \cline{3-9}
\multicolumn{1}{|l|}{}                        & \multicolumn{1}{l|}{}                             & nl  & \multicolumn{1}{l|}{$1.57\pm0.21$}  & \multicolumn{1}{l|}{$\quad 2.72$} & \multicolumn{1}{l|}{$1.00\pm0.01$}    & \multicolumn{1}{l|}{}       &   \multicolumn{1}{l|}{1.55} &  5.5\% &  &  \\ \cline{2-9}
\multicolumn{1}{|l|}{}                        & \multicolumn{1}{l|}{\multirow{2}{*}{$gg+g\kappa(\sigma_8\text{ free})$}} & lin & \multicolumn{1}{l|}{$1.79\substack{+0.22 \\ -0.28}$} & \multicolumn{1}{l|}{$\quad 3.11$}  & \multicolumn{1}{l|}{$1.00\pm0.02$}    & \multicolumn{1}{l|}{$0.80\substack{+0.11 \\ -0.09}$}       &   \multicolumn{1}{l|}{1.10} &  34\%  &  &  \\ \cline{3-9}
\multicolumn{1}{|l|}{}                        & \multicolumn{1}{l|}{}                             & nl  & \multicolumn{1}{l|}{$1.59\substack{+0.29 \\ -0.34}$} & \multicolumn{1}{l|}{$\quad 2.76$}  & \multicolumn{1}{l|}{$1.00\pm0.02$}    & \multicolumn{1}{l|}{$0.80\substack{+0.11 \\ -0.09}$}       &  \multicolumn{1}{l|}{1.46} &  8.8\%  &  &  \\ \cline{1-9}
\smallskip
\end{tabular}
\label{tab:flux_0p4_bg_Dz.}
\end{table*}



\label{lastpage}
\end{document}